\documentclass{iopart}
\usepackage{amssymb}
\usepackage{graphicx}
\usepackage[square,comma,numbers,sort&compress]{natbib}
\usepackage{bm}
\usepackage{mathrsfs}
\usepackage{verbatim}
\usepackage{wasysym}

\def\hf{{\frac{1}{2}}}
\def\d{\mathrm{d}}

\begin{document}

\title{SO(4,1) Yang-Mills theory of quantum gravity}
\author{Timothy D. Andersen}

\address{Dept. of Mathematics, Rensselaer Polytechnic Institute, Troy, NY, 12180\label{addr1}\\\emph{Present Address:} 2 Eaton St., Ste. 500, Hampton, VA 23669}

\date{Received: date / Accepted: date}

\maketitle

\begin{abstract} The search for a quantum theory of gravity has become one of the most well-known problems in theoretical physics.  Problems quantizing general relativity because it is not renormalizable have led to a search for a new theory of gravity that, while still agreeing with measured observations, is renormalizable.  In this paper, a spin-1 Yang-Mills force theory with a SO(4,1) or {\em de Sitter} group symmetry is developed. By deriving the standard geodesic equation and the first post-Newtonian approximation equations, it is shown that this theory, coupled to Dirac fields, predicts all N-body and light observations of gravitational phenomena to within experimental accuracy. Furthermore, because of the separation of gauge covariance from coordinate diffeomorphism, the theory satisfies the strong equivalence principle while maintaining a Minkowski coordinate metric. Cosmology is also briefly addressed: Vacuum energy is the most common explanation for the accelerating expansion of the universe but suffers from the drawback that any reasonable prediction of it is 120 orders of magnitude too large. The de Sitter solution to the Einstein Field Equations is an alternative to vacuum energy as an explanation for the accelerating expansion of the universe but only if the universe is approximately a vacuum. The proposed gauge theory, however, avoids both these problems and, cosmologically, the accelerating expansion of the universe is shown as a consequence of the de Sitter group Lie algebra. In addition, with quantized mass, because it is a generic massless, semi-simple Yang-Mills theory, it is mathematically proved to be a perturbatively renormalizable quantum theory of gravity.
\end{abstract}

\section{Introduction}
That general relativity is able to subsume Newtonian gravity and explain phenomena that do not fit into the Newtonian framework such as Mercury's perihelion precession, light deflection, and gravitational red-shift as well as being compatible with special relativity has brought it wide acceptance.  In recent years, however, as measurement tools have grown more accurate and new observations made, the necessity of introducing either tunable parameters or exotic forms of matter and energy to make general relativity fit those measurements has brought the theory into question.  Observations of galactic rotational curves and gravitational lensing \cite{Zwicky:1937} \cite{Rubin:1970} \cite{Clowe:2006} have demanded the introduction of dark matter and accelerating expansion of the universe \cite{WMAP:2008} dark energy.  At present there is no consensus on what these substances are.

Besides the problems with general relativity at the macroscopic scale, achieving a quantum theory of gravity has become one of the most significant unsolved problems in physics.  Attempts to place the Einstein-Hilbert action,\begin{equation}S_{EH} = \int d^4x\sqrt{-g}R,\end{equation} where $R$ is the Ricci scalar, into a functional path integral ensemble, have all failed, giving nonsense results.  The immediate source of the difficulty is that the theory is not renormalizable creating an infinite number of counterterms in perturbation expansions.  Another, less theoretically troubling but mathematically difficult issue is that the expansion of the Lagrangian fails to terminate because of the volume element $\sqrt{-g}$ and the inverse metric $g^{\mu\nu}$.  Therefore, unlike the actions of other forces, the gravitational action is not finite polynomial.  A disaster also occurs quantizing the weak force but disappears when it is unified with the electromagnetic force, motivating the quest to unify gravity with the other forces in the hope that it will become renormalizable \cite{Zee:2003}. 

Although elegant and surprisingly accurate until recent decades, because the Einstein-Hilbert action causes such deep, unresolvable problems at the quantum level, there is some motivation to find an alternative to general relativity (GR). Despite its current acceptance, the diffeomorphic group of GR was a counterintuitive choice for a theory of gravity given the success of electromagnetism in the 19th century.
As Wald points out \cite{Wald:1984},
\begin{quote}
Maxwell's theory is a remarkably successful theory of electricity, magnetism, and light which is beautifully incorporated into the framework of special relativity.  Therefore, one might expect that the next logical step would have been to develop a new theory of the other classical force, gravitation, which would generalize Newton's theory and make it compatible with special relativity in the same way that Maxwell's theory generalized Coulomb's electrostatics.  However, Einstein chose an entirely different path[.]
\end{quote} 

Although general relativity has stood the test of time for nearly a century, its strong aspects remain untested. All precision experiments (classic tests) of general relativity have been done within the solar system (with the exception of binary pulsar precession) where spherically symmetric, weak gravity prevails and only the first order post-Newtonian Einstein-Infeld-Hoffman (EIH) equations have been confirmed \cite{Misner:1973} \cite{Wald:1984}.  An example of the best recent evidence for strong field general relativity is the periastron precession of the double pulsar system PSR J0737-3039A/B \cite{Kramer:2006}, which is a first post-Newtonian order confirmation.  Furthermore, recent measurements of the cosmic microwave background, where spacetime curvature is most likely to appear, show none at all \cite{Komatsu:2011} with the inflationary theory as a potential explanation \cite{Weinberg:2008}.  

The difference between the EIH equations and the full equations of GR has invited a large number of alternative theories of gravity such as are discussed in \cite{Will:1993}, but virtually all are metric theories and retain the diffeomorphic spacetime picture with the associated problems mentioned above.  Non-metric theories, by contrast, offer the hope of eliminating gravity's problems with quantization at the cost of also eliminating the elegant curved spacetime approach.  It has been suggested that the Standard Model approach, where forces have Yang-Mills actions, may be more suitable for a quantum theory of gravity because of its success at explaining the other forces.  The tetrad or veirbein formulation of the Einstein-Hilbert action, for example, has a structure similar to the Yang-Mills \cite{Zee:2003}.  None of these theories have achieved significant success in making predictions that general relativity is not capable of making nor in resolving, entirely, the problems with the quantum theory.  Thus, there has been, as yet, no motivation for replacing general relativity with any of these nor of accepting any as quantum gravity's true representation. Ideally, such a theory would: (1) agree with all of GR's measured predictions to experimental accuarcy, (2) be a renormalizable quantum theory, and (3) explain some of the mysteries that GR fails to explain such as dark energy.

In this paper, I present a Yang-Mills theory of gravity with a SO(4,1) or de Sitter group that does all of these. This paper represents the first time that a Yang-Mills theory based on this group has been shown to agree with all precise observations of gravitational phenomena. Because a non-Abelian SO(4,1) theory has local Lorentz symmetry separate from coordinate diffeomorphism, the theory complies with the Einstein Equivalence Principle (EEP) and the Strong Equivalence Principle (SEP) (which even most alternate metric theories do not satisfy \cite{Will:1993}). All physical theories are Poincar\'e symmetric, but, given that the Poincar\'e group is not semi-simple, the de Sitter group is a natural choice for a semi-simple extension. The theory uses the Minkowski metric for its coordinate system because the de Sitter group representation is dependent on a coordinate system. This coordinate system, however, is not an active coordinate system as in general relativity.

This theory has a spin-1 rather than spin-2 potential. However, because it has an SO(4,1) group, the group indexes are spin indexes as well as group indexes. This is only possible for the Poincar\'e, de Sitter, and anti-de Sitter groups. Thus, the metric potential, $G_{\mu\nu}$, has two spin indexes with the second one doubling as a group index. It is a pseudotensor. Although it is commonly assumed that gravity must be a spin-2 tensor, there is no physical evidence that this is the case. The physical evidence for gravity only requires a potential with two spin indexes not necessarily that it be a tensor.

Other gauge theoretic approaches to gravitation suffer from problems that the SO(4,1) theory does not. Conformal gravity, another Yang-Mills approach \cite{Riegert:1986}, for example, although having some interesting theoretical properties, does not agree with Newtonian gravitation. To deal with this problem, Newtonian interaction has to be modified at long range, a requirement that is significantly at odds with astronomical observations. Although modified Newtonian gravity has been offered as a potential explanation for dark matter \cite{Mannheim:2006}, there are significant draw backs to any gravity-based (as opposed to particle-based) explanation for dark matter. In particular, dark matter tends to appear in galaxies but often not globular clusters of similar size \cite{Freeman:2006}. Indeed, the high variability of dark matter suggests it is a substance, e.g., Weakly Interacting Massive Particles (WIMPs), and not the result of additional gravitational fields. Other gauge theoretic approaches to gravity exist as well, some of which, unlike conformal gravity and the proposed SO(4,1) theory, violate the equivalence principle. These theories have not gained wide attention in the physics community. The Macdowell--Mansouri action, which is based on the anti-de Sitter group, is an attempt to reconcile macroscopic general relativity with a microscopic de Sitter group, but it is not locally de Sitter invariant. Also, because it is not a Yang-Mills theory and reproduces general relativity in the continuum limit, it is not renormalizable \cite{Wise:2010}.

Yang-Mills theories have a finite, polynomial action with a {\em renormalizable} quantization provided the coupling constant has non-negative mass dimension.  Since $G$ has a mass-dimension of $-2$, a non-dimensional gravitational constant, $a_g$, is needed for quantum predictions. Because the coupling constant is non-dimensional, mass, energy, momentum, and angular momentum are quantized, with mass quantum $m_g$, which has been suggested in the context of other theories of gravity \cite{Wesson:2003}.

The second component to the theory is that gravitational mass is derived as a Yang-Mills current, equivalent to inertial mass, which has a different form than standard stress-energy tensors. In Einstein's theory of gravity, matter couples to gravity through the metric tensor for coordinate space. Based on how, e.g., the Dirac action couples to the metric tensor, it is easy to see that stress-energy tensors of fermions in the rest frame, treated classically, are point-masses. When combining together into stars and planets, these bodies behave like point masses as well. In each rest frame, each of these bodies has only one non-zero component in its stress energy tensor, $T^{00}=m$. In a Yang-Mills theory of gravity, however, particles have space-space components, $J_{ij}\neq 0$.

The reason there is a difference between how mass is defined in General Relativity versus the Yang-Mills theory is in how Lagrangians, particularly the Dirac Lagrangian,
\begin{equation}
\mathcal{L}_{Dirac} = i\bar{\psi}\gamma^\mu\partial_\mu \psi - m\bar{\psi}\psi,
\end{equation} couple to the gravitational potential. The energy-momentum tensor for the Dirac equation (excluding the energy of gravitational fields) is,
\begin{equation}
T^{\mu\nu} = i\bar{\psi}\gamma^\mu\partial^\nu\psi.
\end{equation} In the Yang-Mills theory, the coupling is through the matrix potential, $A_\mu$, rather than a metric, replacing the ordinary derivative with a covariant derivative: $D_\mu = \partial_\mu + im_g A_\mu$. (The value of $m_g$, the mass quantum, is unknown.) The gravitational current is similar to the Dirac electromagnetic current, \begin{equation}J^\mu = \bar{\psi}\gamma^\mu\psi.\end{equation} Thus, for a matter particle's Lagrangian, the gravitational current, \begin{equation}J^{\mu\nu} = \delta \mathcal{L}/\delta G_{\mu\nu} = im_g\bar{\psi}\gamma^\mu P^\nu\psi\end{equation} no longer depends on $\partial^\nu\psi$. The 4 by 4 matrices $P^\nu$ are the de Sitter group generators for spinors for rotations in the $x_\mu$-$x_4$ plane \cite{Parikh:2005},
\begin{equation}
P^0 = \frac{i}{2}\left(\begin{array}{cc} -I & 0 \\ 0 & I\end{array}\right)\quad P^i = \frac{i}{2}\left(\begin{array}{cc} 0 & \sigma_i \\ -\sigma_i & 0\end{array}\right),
\end{equation} where $\sigma_i$ are the Pauli matrices. The generators can also be written $P^0=-\frac{i}{2}\gamma^0$ and $P^i=\frac{i}{2}\gamma^i$. This means that
\begin{equation}
J^{\mu\mu} = -\frac{m_g}{2}\bar{\psi}\psi,
\end{equation} and $J^{\mu\nu}=-J^{\nu\mu}$ for $\mu\neq \nu$. (Note how the flipped sign from that of electromagnetism means that like signed particles attract.) In the case of GR the derivative ensures that the $\psi$ field that is not changing in space has no momentum or kinetic energy. Without the derivative, the spin and internal kinetic energy of the $\psi$ field, which may be constant in space, also have energy. 

The benefit of this approach is that the Higgs mechanism remains unchallenged as the source of inertial mass in the Lagrangian while the spin and internal kinetic energy generate a gravitational mass equal to the inertial mass.

With these definitions, this Yang-Mills theory predicts N-body observations such as pertain to the Solar System and binary pulsars including radiation reaction, all with a finite polynomial, positive definite action that is renormalizable.

The theory offers a cosmology slightly different from the $\Lambda$CDM model. For example, there is a distinction between Doppler and gravitational redshift, i.e. redshift caused by matter moving within space and redshift caused by the expansion of space itself. For the expansion of matter, a linear coasting cosmology is derived.  The linear cosmology has been shown to fit observations such as Type-Ia supernovae and the Cosmic Microwave Background related to the later universe well \cite{Benoit:2008a} \cite{Kaplinghat:2000}. The gravitational potential determines the degree of gravitational redshift which gives the expansion of space itself. The combination of the expansion of matter with the expansion of space leads to the following novel result: while in the Poincar\'e group approximation of the de Sitter theory, the linear coasting model dominates the later universe with the gravitational redshift vanishing, in the de Sitter theory the combination of the Doppler and gravitational redshift predicts a linear acceleration in total observed redshift. 

In general relativity, the de Sitter solution to the Einstein field equations is an {\em ad hoc} explanation for the acceleration of distant galaxies \cite{Weinberg:2008}. For the de Sitter Yang-Mills theory it is anything but ad hoc. It derives automatically from the quantum theory and applies to both the cosmological and quantum realms (as well as everything in between) and provides a dynamical formula for the acceleration. Therefore, it has a compelling theoretical basis for the accelerating expansion on par with vacuum energy but without the problem the vacuum energy theory has with a discrepancy between the observed and predicted cosmological constant of 120 orders of magnitude \cite{Zee:2003}.

The de Sitter model for accelerating expansion in this paper is explained from a particle as well as a geometric point of view by observing that, in the de Sitter Lie algebra, momentum produces spin by the coupling of momentum generators $[V^\mu,V^\nu] = iM^{\mu\nu}$ where $V^\nu$ are momentum generators and $M^{\mu\nu}$ are spin generators. In the Poincar\'e Lie algebra spin can only be produced if spin is already present because $[V^\mu,V^\nu] = 0$. Because the de Sitter theory produces spin from momentum without an initial source of spin, it causes the expansion of the universe to produce gravitons that in turn couple to spin which in turn create gravitons that couple back to momentum. The back and forth production of gravitons leads to an accelerating redshift. This explanation is similar to the explanation for Thomas precession where the orbital motion of an electron increases its spin---an anomaly explained in the early 20th century which can be shown from the Lorentz Lie algebra \cite{Jackson:1999}. Hence, gravitons produce ``dark energy'' because of the relationship between momentum and spin in a de Sitter symmetry group. No cosmological constant is required, and no vacuum energy is modelled. The accelerating expansion effect is produced directly from the Yang-Mills equations. Observations currently attributed to dark matter such as galactic rotational curves and anomalous lensing are not addressed in this paper.

The paper is organized as follows: Sec. \ref{sec:conserved} describes the conserved quantities of mass, linear momentum, and angular momentum in a de Sitter Yang-Mills current; Sec. \ref{sec:yangmills} shows a derivation of the field equations for the SO(4,1) Yang-Mills theory and its formulation as two covariant field equations; Sec. \ref{sec:geo} derives the equation for particle motion and shows that it is the same as the geodesic equation; Sec. \ref{sec:sep} gives a discussion for how the theory satisfies the Strong Equivalence Principle (SEP) which GR also satisfies; Sec. \ref{sec:quantum} briefly mentions the mathematical justification for a renormalizable quantization; Sec. \ref{sec:tests} shows that the theory agrees with measurements within the solar system including redshift, perihelion precession, time dilation and light bending (\ref{sec:traj}), binary pulsar precession (\ref{sec:1pn}), the quadrupole formula for gravitational radiation as well as showing that there are no dipole/monopole moments (\ref{sec:energyloss}), and the equations for an expanding Robertson-Walker universe (\ref{sec:ae}); Sec. \ref{sec:disagreement} explores any possible experimental disagreement with the theory; Sec. \ref{sec:novel} discusses tests that may confirm or deny the theory's novel predictions; and Sec. \ref{sec:relwork} discusses related work.  There are also two Appendices.

\section{The Theoretical Foundation of SO(4,1) Yang-Mills gravity}
Although {\em ad hoc} gravitational theories exist, a good theory tends to be based on a guiding principle. In developing this theory, the guiding principle is that gravity is an ordinary force in the standard, SU(3)$\otimes$SU(2)$\otimes$U(1) model.

\subsection{Conserved quantities in the classical limit from the Dirac action}
\label{sec:conserved}
In this section, we discuss the conserved quantities for the SO(4,1) Yang-Mills theory. Since the majority of matter involved in gravitational observations is either baryon matter or dark matter/energy of unknown composition, we will focus on baryon matter and, specifically, fermions and radiation and neglect, for now, gravitational energy. As discussed in the introduction, the gravitational current for the de Sitter Yang-Mills theory is similar to the electromagnetic current:
\begin{equation}
J^\mu = e\bar{\psi}\gamma^\mu\psi.
\end{equation} The gravitational current is,
\begin{equation}
J^{\mu\nu} = im_g\bar{\psi}\gamma^\mu P^\nu\psi,
\label{eqn:energy}
\end{equation} where $P^0=-\frac{i}{2}\gamma^0$ and $P^i = \frac{i}{2}\gamma^i$. We also have a spin density,
\begin{equation}
S^{\mu\nu\lambda} = i\frac{m_g}{2}\bar{\psi}\gamma^\mu\sigma^{\nu\lambda}\psi,
\label{eqn:spindens}
\end{equation} where $\sigma^{\mu\nu} = \frac{i}{2}[\gamma^\mu,\gamma^\nu]$. 

The derivation of these equations from the Dirac Lagrangian is relatively simple:
\begin{equation}
S_{Dirac} = \int d^4x\,i\bar{\psi}\gamma^\mu(\partial_\mu + im_gA_\mu)\psi,
\end{equation} is the Dirac action and $A_\mu$ is a matrix potential for the De Sitter group with $m_g$ the mass quantum. Break the matrix potential into a sum over de Sitter generators for spinor fields,
\begin{equation}
iA_\mu = G_{\mu\nu}P^\nu + \frac{1}{2}H_{\mu\nu\lambda}\sigma^{\nu\lambda}.
\end{equation} Now apply the Euler-Lagrange equation to $S_{Dirac}$ with respect to $G_{\mu\nu}$ and $H_{\mu\nu\lambda}$, and we arrive at the energy \ref{eqn:energy} and spin density \ref{eqn:spindens}. Because $\gamma^\mu\gamma^\mu = \eta^{\mu\mu}$ and, otherwise, $\gamma^\mu\gamma^\nu=-\gamma^{\mu}\gamma^\nu$ when $\mu\neq\nu$, the energy density simplifies to \cite{Zee:2003},
\begin{equation}
J^{\mu\nu} = -\frac{m_g}{2}\bar{\psi}(\delta^{\mu\nu} - iP^{\mu\nu})\psi,
\end{equation} where $-P^{i0} = P^{0i} = -\sigma^{0i}$ and $P^{ij} = \sigma^{ij}$. The off-diagonals of $J^{\mu\nu}$ represent the contribution of total angular momentum to the energy density. If the off-diagonals are zero, with no angular momentum, then in spherical coordinates (using the coordinate convention in \cite{Misner:1973} $(t,r,\phi,\theta)$)
\begin{equation}
J_{tt} = J_{rr} = -\frac{m_g}{2}\bar{\psi}\psi,\, J_{\phi\phi} = -\frac{m_g}{2}\bar{\psi}\psi r^2\sin^2\theta,\, J_{\theta\theta} = -\frac{m_g}{2}\bar{\psi}\psi r^2.
\end{equation} For a ``point'' mass, which is the main concern in classical N-body dynamics, all the mass is concentrated at $r=0$. Thus,
\begin{equation}
J_{tt} = J_{rr} = -\frac{m_g}{2}\bar{\psi}\psi,\, J_{\phi\phi} = J_{\theta\theta} = 0.
\end{equation} This simplifies, in classical equations, to
\begin{equation}
J_{tt} = J_{rr} = -m.
\label{eqn:sphenergy}
\end{equation}

Only this pseudotensor is conserved because the SO(4,1) symmetry pre-empts the coordinate diffeomorphic symmetry. General relativity assumes diffeomorphism-covariance which implies that the quantity \ref{eqn:energy} is conserved by Noether's theorem.  Kretschmann pointed out to Einstein himself (and Einstein conceded), however, that general covariance, i.e. coordinate covariance, was a fictitious symmetry group generating no physical predictions \cite{Norton:2003}.  Many theories can be made that are generally covariant and predict different physical outcomes.  The physical restrictions on the theory are important and fix the theory.  Diffeomorphism-covariance in general relativity is an active symmetry that is unrestricted.  Imposing an SO(4,1) gauge symmetry on spacetime implies that, once it is fixed, the gauge prefers a particular coordinate system.  With a preferred set of reference frames, although the coordinate system can be changed at will, the theory is not actively diffeomorphism-covariant.  Hence, Noether's theorem does not apply to the metric, and the stress-energy-momentum tensor of General Relativity is not a conserved quantity.  This is not the same as having a preferred coordinate system overall, only that fixing the gauge destroys the diffeomorphic freedom that general relativity assumes.

\subsection{Yang-Mills theory and derivation of the field equations}
\label{sec:yangmills}
In this section, I begin with standard Yang-Mills theory and derive a locally de Sitter theory of quantum gravity using the de Sitter Lie algebra. Yang-Mills theory describes forces as exchanges of gauge bosons based on a group symmetry.  The Standard Model of quantum field theory currently contains three forces, electromagnetism, the weak force, and the strong force in a U(1)$\otimes$SU(2)$\otimes$SU(3) group symmetry, all using the Lie algebras of semi-simple groups.

A conventional SU(N) Yang-Mills theory has action,
\begin{equation}
S = -\frac{1}{4g^2}\int \d^4 x\, F_{\mu\nu}F^{\mu\nu},
\end{equation} where $g$ is the coupling constant and
\begin{equation}
F_{\mu\nu} = \partial_\mu A_\nu - \partial_\nu A_\mu - i[A_\mu,A_\nu],
\end{equation} is the ``force'' (a misnomer carried over from the electromagnetic Lorentz force) where $A_\mu$ is a matrix potential from the SU(N) group.  For the SO(4,1) group these are $5\times 5$ matrices. (Note that in the previous section we absorbed a factor of $i$ into $A_\nu$ to make the matrices real.) The theory is invariant under a gauge transformation $U(x)$ such that $A'_\mu = U(x)A_\mu U^{\dagger}(x) - i/g\partial_\mu U(x) U^{\dagger}(x)$. For infinitesimal transformations, $U(x) = e^{i\chi(x)}\approx I + i\chi(x)$, $A'_\mu = A_\mu + \partial_\mu \chi(x) + i[\chi(x),A_\mu]$ is the gauge transformation.

Yang-Mills equations, expressed in group component form, have the action,
\begin{equation}
S = -\frac{1}{4g^2}\int \d^4x\, F_{\mu\nu}^aF^{\mu\nu a},
\label{eqn:ymaction}
\end{equation}
\begin{equation}
F_{\mu\nu}^a = \partial_\mu A_\nu^a - \partial_\nu A_\mu^a + f^{abc}A_\mu^bA_\nu^c,
\end{equation} and $f^{abc}$ is the group symbol.

The SO(4,1) potential is written as a sum of generators of Lorentz rotations and boosts and of boost/rotations with respect to $x_4$,
\begin{equation}
A_{\mu} = G_{\mu\nu}V^\nu + \hf H_{\mu\nu\lambda}M^{\nu\lambda},
\end{equation} where the pseudotensor potential $G_{\mu\nu}$ couples to $J^{\mu\nu}$ and pseudotensor $H_{\mu\nu\lambda}$ to $S^{\mu\nu\lambda}$.  In this sense, the theory resembles Einstein-Cartan theory but not in the same way as the Macdowell-Mansouri's action which is neither de Sitter invariant nor renormalizable \cite{Wise:2010}.

Following the de Sitter algebra of Appendix \ref{sec:lorentz},  the ``forces'' of gravity decompose into a covariant form with two equations,
\begin{eqnarray}
 E_{\mu\nu\lambda} = \partial_\mu G_{\nu\lambda} - \partial_\nu G_{\mu\lambda} +\nonumber\\ \eta^{\sigma\rho}\left(G_{\mu\lambda}H_{\nu\sigma\rho} - G_{\mu\rho}H_{\nu\sigma\lambda} - G_{\nu\lambda}H_{\mu\sigma\rho} + G_{\nu\rho}H_{\mu\sigma\lambda}\right),
\label{eqn:transforce}
\end{eqnarray}
and
\begin{eqnarray}
F_{\mu\nu\alpha\beta} = \partial_\mu H_{\nu\alpha\beta} - \partial_\nu H_{\mu\alpha\beta} + G_{\mu\alpha}G_{\nu\beta} - G_{\mu\beta}G_{\nu\alpha} + \Phi_{\mu\nu\alpha\beta} - \Phi_{\mu\nu\beta\alpha}
\label{eqn:rotforce}
\end{eqnarray}  where
\begin{eqnarray}
\Phi_{\mu\nu\alpha\beta}  = \hf\eta^{\sigma\rho}\left(H_{\mu\sigma\alpha}H_{\nu\rho\beta} - H_{\mu\sigma\alpha}H_{\nu\beta\rho} -\right.\nonumber\\ \left.H_{\mu\alpha\sigma}H_{\nu\rho\beta} + H_{\mu\alpha\sigma}H_{\nu\beta\rho}\right).
\end{eqnarray} Thus, the action of gravity is,
\begin{equation}
S_{gravity} = \frac{1}{4a_g}\int \d^4x\, E_{\mu\nu\lambda} E^{\mu\nu \lambda} + \hf F_{\mu\nu\alpha\beta} F^{\mu\nu\alpha\beta}.
\end{equation}  Here the coupling constant is $a_g=g^2$.

The classical equations of motion of any Yang-Mills theory can be found by the variation of the action, $\delta S=0$, via the Euler-Lagrange equations,
\begin{equation}
\partial^\mu\left(\frac{\partial \mathcal{L}}{\partial (\partial^\mu A_\nu{}^a)}\right) - \frac{\partial\mathcal{L}}{\partial A_\nu{}^a} = 0,
\end{equation} where $\mathcal{L} = F_{\mu\nu}^aF^{\mu\nu}_a$ is the Lagrangian.  Evaluating the Euler-Lagrange equations for the Yang-Mills Lagrangian, the equations of motion are,
\begin{equation}
\partial^\mu F_{\mu\nu} - i[F_{\mu\nu},A^{\mu}] = -8\pi g^2 J_\nu,
\label{eqn:ymfeq}
\end{equation} where $J_\nu$ is the conserved current.

Applying the definition of the group symbol again and the previous definitions \ref{eqn:transforce} and \ref{eqn:rotforce} to the field equations (\ref{eqn:ymfeq}) yields the following equations, analogous to the Einstein field equations and the Cartan torsion equations respectively:
\begin{eqnarray}
\label{eqn:motion1}
 \partial^\mu E_{\mu\nu\lambda} + \eta^{\sigma\rho}\left(E_{\mu\nu\lambda}H^\mu{}_{\sigma\rho} - E_{\mu\nu\rho}H^\mu{}_{\sigma\lambda} -\right.\nonumber\\\left. G^\mu{}_{\lambda}F_{\mu\nu\sigma\rho} + G^\mu{}_{\rho}F_{\mu\nu\sigma\lambda}\right) & = & -8\pi a_gJ_{\nu\lambda}\\
\label{eqn:motion2}
\partial^\mu F_{\mu\nu\alpha\beta} + E_{\mu\nu\alpha}G^\mu{}_{\beta} - E_{\mu\nu\beta}G^\mu{}_{\alpha} +\Sigma_{\nu\alpha\beta} - \Sigma_{\nu\beta\alpha} & = & -8\pi a_gS_{\nu\alpha\beta}
\end{eqnarray}  where
\begin{eqnarray}
\Sigma_{\nu\alpha\beta} = \hf\eta^{\sigma\rho}\left(F_{\mu\nu\sigma\alpha}H^\mu{}_{\rho\beta} - F_{\mu\nu\sigma\beta}H^\mu{}_{\beta\rho} - \right.\nonumber\\\left.F_{\mu\nu\alpha\sigma}H^\mu{}_{\rho\beta} + F_{\mu\nu\alpha\sigma}H^\mu{}_{\beta\rho}\right).
\end{eqnarray} The coupling constant, $a_g=g^2$, is non-dimensional, but, because the size of the mass quantum, $m_g$, is not known, the standard coupling constant, $\kappa = G/c^4$, in units such that $\kappa=1$, is sufficient for the classical equations.

Boundary conditions are: 
\begin{equation}
G_{\mu\nu}\rightarrow\eta_{\mu\nu},\quad \partial_\lambda G_{\mu\nu} \rightarrow 0
\end{equation} and 
\begin{equation}
H_{\mu\nu\lambda}\rightarrow 0,\quad \partial_\alpha H_{\mu\nu\lambda} \rightarrow 0
\end{equation} as $x_\mu\rightarrow\infty$.

Gauge transformations can be done in component form. Consider the $SO(4,1)$ de Sitter transformation $\Delta(x) = \xi_\mu(x) V^\mu + \hf \chi_{\mu\nu} M^{\mu\nu}$. The gauge transformation is 
\begin{equation}
G'_{\mu\nu} = G_{\mu\nu} + \partial_\mu \xi_\nu(x) + \eta^{\rho\lambda}(\xi_\lambda(x)H_{\mu\rho\nu} - \xi_\lambda(x)H_{\mu\nu\kappa} + \chi_{\lambda\nu}(x)G_{\mu\rho} - \chi_{\nu\rho}G_{\mu\lambda})
\end{equation} and
\begin{equation}
H'_{\mu\nu\lambda} = H_{\mu\nu\lambda} + \partial_\mu \chi_{\nu\lambda}(x) + \xi_\nu(x)G_{\mu\lambda} + \hf\eta^{\kappa\rho}(\chi_{\kappa\nu}(x)H_{\mu\rho\lambda} - \chi_{\kappa\nu}(x)H_{\mu\lambda\rho} - \chi_{\nu\kappa}(x)H_{\mu\rho\lambda} + \chi_{\nu\kappa}(x)H_{\mu\lambda\rho}),
\end{equation} which can be found from the Lie algebra.

\subsection{Equations for evolution of matter}
\label{sec:covform}
The continuity equations determine the evolution of matter. By Noether's theorem, for any matter action, $\mathcal{S}_M$, e.g., Dirac's, the relations between the potential $G_{\mu\nu}$ and the stress-energy-tensor $J^{\mu\nu}$ and the potential $H_{\mu\nu\lambda}$ and the spin-density $S^{\mu\nu\lambda}$ are,
\begin{eqnarray}
J^{\mu\nu} & = & \frac{\delta \mathcal{S}_M}{\delta (G_{\mu\nu})},\\
S^{\mu\nu\lambda} & = & \frac{\delta \mathcal{S}_M}{\delta ( H_{\mu\nu\lambda})},
\end{eqnarray} and the continuity equations are,
\begin{eqnarray}
\label{eqn:cont}
D^\mu J_{\mu\nu} = 0,\\
\label{eqn:conts}
D^\mu S_{\mu\nu\lambda} = 0,
\end{eqnarray} \cite{Joshi:1985}, where, for a generic source $J_\mu{}^a$ and potential $A_\mu{}^a$, $D_\mu J_\nu = \partial_\mu J_\nu - i[A_\mu,J_\nu]$ is the covariant derivative.  Applying the de Sitter Lie algebra to \ref{eqn:cont} and \ref{eqn:conts}, the full continuity equations are,
\begin{eqnarray}
\label{eqn:cont2}
\partial^\mu J_{\mu\lambda} + \eta^{\sigma\rho}\left(G^\mu{}_{\lambda}S_{\mu\sigma\rho} - G^\mu{}_{\rho}S_{\mu\sigma\lambda} -\right.\nonumber\\\left. J^\mu{}_{\lambda}H_{\mu\sigma\rho} + J^\mu{}_{\rho}H_{\mu\sigma\lambda}\right) = 0,\\
\partial_\mu S^{\mu\alpha\beta} + G^\mu{}_{\alpha}J_{\mu\beta} - G^\mu{}_{\beta}J_{\mu\alpha} +  \Pi^{\alpha\beta} - \Pi^{\beta\alpha} = 0
\end{eqnarray} where
\begin{eqnarray}
\Pi_{\alpha\beta} = \hf\eta^{\sigma\rho}\left(H^\mu{}_{\sigma\alpha}S_{\mu\rho\beta} - H^\mu{}_{\sigma\beta}S_{\mu\beta\rho} -\right.\nonumber\\\left. H^\mu{}_{\alpha\sigma}S_{\mu\rho\beta} + H^\mu{}_{\alpha\sigma}S_{\mu\beta\rho}\right).
\end{eqnarray}

There are nonlinearities in the field equations \ref{eqn:motion1}-\ref{eqn:motion2} and the continuity equations because gravitons have spin, polarization and momentum and couple to themselves.  These nonlinearities are too small to detect except at the cosmological scale (Sec. \ref{sec:ae}) and close to strongly gravitating objects moving at high speeds or rates of rotation.

When $H_{\mu\nu\lambda} = 0$, the field equations without an angular momentum source revert to the Abelian equation,
\begin{equation}
\label{eqn:abelian}
\partial^\mu (\partial_\mu G_{\nu\lambda} - \partial_\nu G_{\mu\lambda}) = -8\pi a_g J_{\nu\lambda},
\end{equation} which, under the harmonic gauge condition, is,
\begin{equation}
\label{eqn:relaxed}
\square G_{\mu\nu} = -8\pi a_g J_{\mu\nu},
\end{equation} which relates $J_{\mu\nu}$ to the pseudotensor $G_{\mu\nu}$. This equation is most useful for experimental predictions since $H_{\mu\nu\lambda}$ is usually too small to detect. In the following, I will use the known gravitational constant in Planck units, $G/c^4 = 1$, for classical equations assuming that the mass quantum, $m_g$, is negligible for N-body dynamics and cosmology.

\subsection{Equation of test particle motion}
\label{sec:geo}
There are two forms of test particle: (1) photons and (2) massive particles. 

\subsubsection{The geodesic of light and other massless particles}
A test particle with null geodesic and no internal structure is represented as a parameterized path $x^\mu(\tau)$. $G_{\mu\nu}$, takes the place of the metric normally found in Lagrangians for test particle motion. Indexes are raised and lowered with the Minkowski metric $\eta_{\mu\nu}$. The Lagrangian for motion is:
\begin{equation}
\mathcal{L} = G_{\mu\nu}\frac{dx^\mu}{d\sigma}\frac{dx^\nu}{d\sigma} + H_{\mu\nu\lambda}\frac{dx^\mu}{d\sigma}\frac{dx^\nu}{d\sigma}\frac{dx^\lambda}{d\sigma}.
\end{equation} This describes both the curvature and twisting of the path as it moves through spacetime along its shortest path.

No observations of gravitational phenomena, save perhaps cosmological observations, are sensitive enough to measure torsion. Therefore, we concern ourselves in this section only with geodesic motion. Let $H_{\mu\nu\lambda}=0$. (A similar assumption is made in applying Einstein-Cartan theory.) Let the Lagrangian for particle motion be,
\begin{equation}
\mathcal{L} = G_{\mu\nu}\frac{dx^\mu}{d\sigma}\frac{dx^\nu}{d\sigma}.
\label{eqn:lagr}
\end{equation}

The equations of motion are given by the Euler-Lagrange equation,
\begin{equation}
\frac{d}{d\sigma}\left(\frac{\partial \mathcal{L}}{\partial (dx_\nu/d\sigma)}\right) - \frac{\partial\mathcal{L}}{\partial x_\nu} = 0.
\end{equation} The equation of motion for the particle (the ``geodesic'') is,
\begin{equation}
\frac{d^2x^\lambda}{d\tau^2} + \frac{1}{2}(G^{-1})^{\lambda\nu}\left[\partial_\rho G_{\mu\nu} + \partial_\mu G_{\rho\nu} -  \partial_\nu G_{\mu\rho}\right]\frac{dx^\mu}{d\tau}\frac{dx^\rho}{d\tau} = 0,
\label{eqn:geo}
\end{equation} which is the geodesic equation. Note: there is a distinction between raising indexes and inverting $G_{\mu\nu}$. Its inverse is $(G^{-1})^{\mu\nu}$ such that $(G^{-1})^{\mu\alpha}G_{\alpha\nu} = \delta^\mu_\nu$, not $G^{\mu\nu} = \eta^{\mu\alpha}\eta^{\nu\beta}G_{\alpha\beta}$.

\subsubsection{The geodesic of massive bodies}
For a test particle with a rest frame, let the Lagrangian for particle motion be,
\begin{equation}
\mathcal{L} = -\frac{1}{2m}G_{\mu\nu} J^{\mu \nu},
\label{eqn:lagr2}
\end{equation} where $J^{\mu \nu}$ is the particle's current given its energy as defined in the section on conserved currents (Sec. \ref{sec:conserved}).

Let the rest frame of the particle be $\bar{x}^\mu = (\sigma,\bar{r},\bar{\theta},\bar{\phi})$ where the particle is at $(\sigma,0,0,0)$. (Both the gauge and the coordinate system must be changed to the rest frame.)  Let $\bar{G}_{\mu\nu}$ be the potential and $\bar{J}_{\mu\nu}$, the current in frame $\bar{x}^\mu$.  From \ref{eqn:sphenergy} the rest frame current is $\bar{J}_{00}=\bar{J}_{\bar{r}\bar{r}}=-m$; therefore, \ref{eqn:lagr2} is,
\begin{equation}
\mathcal{L} = \hf (\bar{G}_{\sigma\sigma} + \bar{G}_{\bar{r}\bar{r}}).
\end{equation}

At the particle's location, $\bar{r}=0$, the metric degenerates into a 2-D metric, $\bar{g}_{\mu\nu} = (-1,1,0,0)$.  Let the potential be traceless, i.e. $\bar{g}_{\mu\nu}\bar{G}^{\mu\nu}=0$, $\bar{G}_{\sigma\sigma}=\bar{G}_{\bar{r}\bar{r}}$.

Replacing $\bar{G}_{\bar{r}\bar{r}}$ with $\bar{G}_{\sigma\sigma}$ in the Lagrangian, we have,
\begin{equation}
\mathcal{L} = \bar{G}_{\sigma\sigma},
\end{equation} or, in the observer's frame,
\begin{equation}
\mathcal{L} = G_{\mu\nu}\frac{dx^\mu}{d\sigma}\frac{dx^\nu}{d\sigma},
\label{eqn:lagr3}
\end{equation} which is the same as \ref{eqn:lagr}. Thus, in both test particle cases the geodesic equation is identical to that of GR.

\subsection{Strong Equivalence Principle}
\label{sec:sep}
Equivalence principles are defining features of gravitational theories.  Newton's theory included the Weak Equivalence Principle (WEP) where an object's weight is proportional to its mass.  Einstein developed the equivalence principle named for him, the Einstein Equivalence Principle (EEP), which states that local experiments in free fall are independent of position and velocity.  When strongly gravitating bodies are taken into account, the validity of the equivalence principles come into doubt because internal structure violates WEP by, e.g., the (never observed) Nordtvedt effect \cite{Will:1993}.  

The Strong Equivalence Principle (SEP) is a natural extension of early equivalence principles.  It has three conditions: (a) an object's weight is proportional to its mass (WEP) for self-gravitating bodies as well as test bodies, (b) the outcome of any local experiment is independent of the velocity of the (freely falling) apparatus, and (c) the outcome of any local experiment is independent of where and when in the universe it is performed \cite{Will:1993}.  

Although a rigorous proof has not been found, GR is the only known metric theory that appears to satisfy SEP strictly.  Because experiment has never found a violation of SEP, any new gravitational theory must satisfy strict constraints on SEP violations.  In the following, the Yang-Mills theory's gauge symmetry allows the required transformations to satisfy SEP.

The Yang-Mills theory satisfies the first condition of SEP, (a), by the arguments of Section 20.6 of \cite{Misner:1973} also found in \cite{Will:2006}:  If the field of a self-gravitating body asymptotically approaches ``flatness'' ($G_{\mu\nu}(R)=\eta_{\mu\nu}$) at some distance, $R$, considered to be the boundary of the local system, this is sufficient to guarantee that a body's self-gravitation and other internal structure does not affect its motion.  Because it is always possible to find a de Sitter gauge that eliminates the gravitational field at the boundary between a local, spherically symmetric, compact system and the external environment, a spherically symmetric self-gravitating body, even a neutron star or black hole, can be regarded as a point particle.  In bimetric theories, which are diffeomorphism-covariant, changing the gauge to eliminate the field at the boundary changes the Minkowski metric, so a coordinate system cannot be found that satisfies this requirement.  With the Yang-Mills theory, however, the gauge can be changed without changing the coordinate system.  It is simple to find a gauge such that $G_{\mu\nu}(R)=\eta_{\mu\nu}$ for a spherically symmetric solution (see \ref{sec:static}).  Therefore, the Yang-Mills theory satisfies condition (a).

The Yang-Mills theory also satisfies the second condition of SEP, (b), by local Lorentz covariance (a subset of local de Sitter covariance) which means that experimental outcomes are independent (by gauge covariance) of Lorentz boosts and rotations. It also satisfies the third condition of SEP, (c), by having no preferred location/time in the field equations.  Unlike Rosen's bimetric theory which has a local gravitational constant that depends on the field \cite{Will:1993}, the Yang-Mills theory has a non-location specific gravitational constant, $a_g$.  Be these arguments, all three conditions of SEP are met.

\subsection{Quantization}
\label{sec:quantum}

The quantization of the theory is given by the generating functional:

\begin{equation}
Z[J,S] = \int DG DH \exp\left[-\frac{i}{4}\int d^4x\, E_{\mu\nu\lambda} E^{\mu\nu \lambda} + F_{\mu\nu\alpha\beta} F^{\mu\nu\alpha\beta} + ia_g\int d^4x G_{\mu\nu}J^{\mu\nu} + H_{\mu\nu\lambda}S^{\mu\nu\lambda}\right]
\end{equation}

Because it is a massless gauge boson Yang-Mills theory, the theory of gravity given in this paper has a finite polynomial, positive definite action (\ref{eqn:ymaction}).  Like all Yang-Mills theories on semi-simple groups the theory is renormalizable \cite{tHooft:2005}.

The standard quantization scheme is appropriate with Fadeev-Popov ghost fields used to derive Feynman rules including the $G_{\mu\nu}$ graviton propagator \cite{Zee:2003}:
\begin{equation}
D^{\lambda\rho}_{\mu\nu}(p) = \frac{-i\delta^{\lambda\rho}}{p^2 + i\epsilon}\left[\eta_{\mu\nu} - \frac{(1 - \xi)p_\mu p_\nu}{p^2 + i\epsilon}\right].
\end{equation} Other than proving renormalizability, which was achieved in the 1970's, any quantum predictions are left for future work.

\section{Experiments and Observations}
\label{sec:tests}
Numerous observations, starting with light bending in 1919, have been made to attempt to confirm predictions of general relativity.  These include gravitational time dilation, redshift, and light bending all of which have been measured within the solar system. Perihelion precession of the planets and binary pulsar precession have also been measured to good first order accuracy \cite{Wald:1984}.  Other observations do not agree with the original theory and have required modifications to the Einstein field equations.  In the following, are derived (1) a static, spherically symmetric solution to the field equations, (2) the 1PN equations of motion for a binary system, (3) the radiation reaction for binary pulsar inspiral, and (4) the accelerating expansion of the universe via a homogeneous, isotropic model.

\subsection{Spherically symmetric solution}
\label{sec:static}
The static, spherically symmetric Schwarzchild solution is one of the most important solutions to the Einstein field equations.  The solution to the de Sitter abelian field equations \ref{eqn:relaxed} is identical up to linear order.  Let spacetime be covered by spherical coordinates $(t,r,\phi,\theta)$ where $\theta$ is colatitude and let the metric be the flat spacetime metric \cite{Wald:1984},
\begin{equation}
ds^2 = -dt^2 + dr^2 + r^2d\theta^2 + r^2\sin^2(\theta)d\phi^2.
\end{equation}  Choosing a harmonic gauge the equations for the potentials can be found by the general Green's function solution \cite{Will:2006} to the relaxed field equations \ref{eqn:relaxed},
\begin{eqnarray}
h_{\mu\nu}(t,{\vec x}) & = & -\int_{\mathcal{C}} \d^3 x' \frac{J_{\mu\nu}(t - |{\vec x} - {\vec x}'|,{\vec x}')}{|{\vec x} - {\vec x}'|},
\end{eqnarray} where $h_{\mu\nu} = (G_{\mu\nu} - \eta_{\mu\nu})/2$ and $\mathcal{C}$ is the past lightcone. Given a form for the stress-energy tensor, $J_{00} = M\delta(r)$ and $J_{rr} = M\delta(r)$,
the solutions to the integral equation (in spherical coordinates) give the potential pseudotensor:
\begin{equation}
G_{00} = -\left(1 - \frac{2M}{r}\right),\,
G_{rr} = 1 + \frac{2M}{r},\,
G_{\theta\theta} = r^2,\,
G_{\phi\phi} = r^2\sin^2(\theta),
\label{eqn:sssmetric}
\end{equation} with the boundary conditions such that $G_{\mu\nu}\rightarrow \eta_{\mu\nu}$ as $r\rightarrow\infty$.  This is also the linearized Schwarzchild solution to the Einstein equations \cite{Wald:1984}.

\subsection{Particle trajectories}
\label{sec:traj}
Particle trajectories in a static, spherically symmetric potential field are identical to those of general relativity.  The only difference is in the radial potential.  The Schwarzchild radial potential $g_{rr} = (1 - 2M/r)^{-1}$ differs from the radial potential derived in Section \ref{sec:static} to quadratic order,
\begin{equation}
G_{rr} = 1 + 2M/r \approx (1 - 2M/r)^{-1} + O((M/r)^2).
\end{equation}   

No tests within the Solar System, whether with respect to the Sun, Earth, or another body (the most precise tests have been done near the Earth) have achieved better than linear order in Schwarzschild coordinates \cite{Misner:1973}\cite{Wald:1984} because of the weak fields involved.  If $M=Gm/c^2$ is the Schwarzchild radius and $m$ is the mass in other units (e.g., kilograms) at the surface of the Sun, $2M_{\astrosun} /r\approx 4.25\times 10^{-6}$ and at the surface of the Earth $2M_{\earth} /r\approx 1.4\times 10^{-9}$.  Both are much too small for higher order effects to be measured, and imperfections in density (e.g., mountains) would make measurements difficult to verify.

Predictions of gravitational time dilation and redshift, meanwhile, are mathematically identical in both theories (to any order) given a static, spherically symmetric field with redshift proportional to $1 + 2M/r$.  Recent high precision measurements of redshift using cesium atoms in a laboratory agree with predictions of both theories \cite{Muller:2010}. Measurements of the geodetic effect and frame dragging by Gravity Probe B are likewise first order accuracy measurements \cite{Everitt:2011}.  Therefore, no tests done within the Solar system to date disprove either theory.  For strong field tests, require looking outside the solar system to binary pulsars and, further on, cosmology where the theories have the best chance of being tested.

\subsection{Post-Newtonian Equations of Motion}
\label{sec:1pn}
The binary pulsar system B1913+16 discovered in 1974 provided one of the first tests of strong field general relativity \cite{Will:2006}.  Because the regular radio pulses of the star, orbiting a relatively inert body, possibly a dead pulsar, allowed precise measurements of the pulsar's orbit, the small deviations of the two body orbit from Kepler's laws can be measured, including orbital damping caused by gravitational radiation, which is a higher order effect than orbital precession \cite{Will:2006}.  Observations of this system over the past thirty-five years have increased the accuracy of the measurements.  Recently, the double binary pulsar system PSR J0737-3039A/B with two active pulsars orbiting each other has provided even greater precision \cite{Kramer:2006}.  The following demonstrates that the field equations of Sec. \ref{sec:yangmills} agree with the observations of these pulsar systems.

The parameterized post-Newtonian (PPN) equations of motion of gravity are the primary vehicle by which not only different theories of gravity (including GR) are compared but provide the equations for predicting the motion of two or more gravitating bodies.  Tests of these equations include the measurement of planetary motion within the Solar system for weak fields and binary pulsar precession and gravitational radiation for stronger fields \cite{Misner:1973} \cite{Will:1993}.  

For a binary pulsar system the GR metric in the post-Newtonian coordinate system for general relativity is (\cite{Will:1993}, 11.52)
\begin{eqnarray}
g_{00} & = & -1 + 2\sum_{a=1,2} m_a/|{\vec x} - {\vec x}_a(t)| + O(\epsilon^4),\\
g_{0j} & = & O(\epsilon^3),\\
g_{ij} & = & \delta_{ij}\left(1 + 2\sum_{a=1,2} m_a/|{\vec x} - {\vec x}_a(t)|\right) + O(\epsilon^4),
\end{eqnarray} where $\epsilon$ is a small parameter such that $v/c\sim \epsilon$ and $GM/c^2r\sim\epsilon^2$ with $m_1$ and $m_2$ being the masses of the bodies and ${\vec x}_1$ and ${\vec x}_2$ their positions in the appropriate harmonic coordinate system.  Because GR satisfies SEP the self-gravitation of the two bodies along with their motion can be included in their masses and they can be modelled as test bodies with the geodesic equation. 

The same method used to derive the GR metric can be used on the relaxed field equations of the Yang-Mills theory (\ref{eqn:relaxed}). The calculations are somewhat involved and are included in Appendix \ref{sec:appendix}. They result in the following potential:
\begin{equation}
G_{00} = -1 + 2\sum_{a=1,2} \frac{m_a}{|{\vec x} - {\vec x}_a|} + O(\epsilon^4),
\end{equation}
\begin{equation}
G_{0j} = O(\epsilon^3),
\end{equation} and
\begin{equation}
G_{ij} = \delta_{ij}\left(1 + 2\sum_{a=1,2}\frac{m_a}{|{\vec x} - {\vec x}_a|}\right) + O(\epsilon^4).
\end{equation}  The velocities of the bodies are $\vec{v}_1=(v_{11},v_{12},v_{13})$ and $\vec{v}_2=(v_{21},v_{22},v_{23})$ and $v_a=\|\vec{v}_a\|$.  (See \ref{sec:appendix} for a derivation of these potentials.)

Since the theory satisfies SEP, for any set of compact nearly-spherical bodies where tidal forces may be neglected, the geodesic equation \ref{eqn:geo} predicts orbital motion.  The post-Keplerian parameters of periastron advance, $\langle\dot{\omega}\rangle$, time delay, $\gamma'$, and Shapiro delay parameters, $r$ and $s$, are also, consequently, the same in both theories. Higher order terms in the PPN equations do not match, however, (see \ref{sec:appendix}) and offer the best chance of ruling out one of the theories with continued observations of binary pulsars.

The final post-Keplerian parameter, orbital speed-up caused by gravitational radiation, $\dot{P_b}$, derives from the quadrupole formula from the Yang-Mills field equations discussed in the following section.

\subsubsection{Orbital speed-up of a binary pulsar system}
\label{sec:energyloss}
Gravitational radiation was first addressed by Einstein shortly after the publication of general relativity, where he and others demonstrated that the primary radiation is quadrupolar in contrast to the dipolar radiation from electromagnetic sources \cite{Einstein:1918}.  The energy loss of a binary system of compact stars caused by radiation was first demonstrated in a paper by Peters and Matthews \cite{Peters:1963}, who derived the energy loss of binary stars in Keplerian orbit leading to the formula for the orbital speed-up,
\begin{eqnarray}
\dot{P}_b =  -\frac{192\pi G^{5/3}}{5c^5}\left(\frac{P_b}{2\pi}\right)^{-5/3}(1-e^2)^{-7/2}\times\nonumber\\ \left(1 + \frac{73}{24}e^2 + \frac{37}{96}e^4\right)m_p m_c(m_p + m_c)^{-1/3},
\label{eqn:speedup}
\end{eqnarray}  with $m_p$ and $m_c$ the masses of the pulsar and companion body respectively and $e$ the orbital eccentricity \cite{Weisberg:2005}.  Although this equation was developed for Keplerian orbits, it applies to post-Keplerian orbits since the only requirement is that the orbit be elliptical, and, although much higher order derivations have been made \cite{Will:1993}, only this equation has been tested.  The equation derives from a multipole expansion of the wave equations for general relativity, and a detailed discussion of multipole expansions for gravitational radiation in general relativity can be found in \cite{Thorne:1980}. For measured radiation-reaction only the lowest order quadrupole term is relevant. The reaction which is what is measured is caused by the conservation of energy which requires a system expelling radiation to slow in some fashion. Orbital speed-up is a reaction to the loss of gravitational potential energy.

Orbital speed-up is a function of energy loss, i.e. radiation-reaction, which to lowest multipole order in General Relativity is (\cite{Misner:1973}, 36.31),
\begin{equation}
\frac{dE}{dt} = -\frac{1}{5}\langle \frac{d^3Q_{jk}}{dt^3}\frac{d^3Q_{jk}}{dt^3}\rangle,
\label{eqn:power}
\end{equation} where $Q_{\mu\nu}$ is the reduced quadrupole moment (the trace free part of the second moment of the mass distribution) such that $Q_{\mu\nu} = q_{\mu\nu} - \frac{1}{3}\delta_{\mu\nu} q$,
\begin{equation}
q_{\mu\nu} = \int \d^3x\, \rho_0x_\mu x_\nu,
\end{equation} and $q = q_\mu{}^\mu$.   The Peters-Matthews formula for energy loss of a binary system (\cite{Will:1993}, 10.80),
\begin{equation}
\frac{dE}{dt} = -\left\langle \frac{\mu^2m^2}{r^4}\frac{8}{15}\left(12v^2 - 11\dot{r}^2\right)\right\rangle,
\label{eqn:pmenloss}
\end{equation} where $m = m_1 + m_2$, $\mu = m_1m_2/m$, $v = |{\vec v}_1 - {\vec v}_2|$, and $r = |{\vec x}_1 - {\vec x}_2|$, can be found from the linearized vacuum equations of general relativity, $\square \bar{h}^{\mu\nu} = 0$, \cite{Peters:1963}\cite{Misner:1973} by the quadrupole relation,
\begin{equation}
\bar{h}_{jk}(t,{\vec x}) = \frac{2}{r}\ddot{q}_{jk}(t - r),
\label{eqn:grwavepot}
\end{equation} for the spatial part of the radiation field (\cite{Misner:1973}, 36.50) where $r$ is the distance from the source.

Peters and Matthews derives the energy loss using the linearized Einstein vacuum equations:
\begin{equation}
g_{\mu\nu} = \eta_{\mu\nu} + h_{\mu\nu},
\end{equation} such that
\begin{equation}
\square \bar{h}_{\mu\nu} = 0,
\end{equation} where $\bar{h}_{\mu\nu} = h_{\mu\nu} - \hf \eta_{\mu\nu}h_{\lambda}{}^{\lambda}$.  

The radiation formula can be found by a plane wave solution to the linearized equations. The linearized Yang-Mills equations for an unpolarized, spinless source,
\begin{equation}
\square G_{\mu\nu} = 8\pi J_{\mu\nu},
\label{eqn:ling}
\end{equation} is the same as the linearized equation for GR. 

Let $G_{\mu\nu} - \eta_{\mu\nu} = h'_{\mu\nu}$. A plane wave solution to the wave equation,
\begin{equation}
\square h'_{\mu\nu} = 0,
\end{equation} is
\begin{equation}
h'_{\mu\nu} = a e_{\mu\nu} \cos(\omega t - \vec{k}\cdot\vec{x}),
\label{eqn:planewave}
\end{equation} where $a$ is the amplitude, $e_{\mu\nu}$ is a symmetric, traceless, transverse, and unitary polarization tensor (as defined in \cite{Peters:1963}), and $\omega$ is the frequency. 

The solution can be expanded by multipoles folllowing the outline in \cite{Will:1993}, Chapter 10. Several steps are left out here but have been worked out there.  

The multipole expansion of the plane wave solution (\ref{eqn:planewave}) is,
\begin{equation}
h'_{\mu\nu} = -4r^{-1}\sum_{m=0}^\infty (1/m!)(\partial/\partial t)^m\int d^3x' J_{\mu\nu}(t - r,\vec{x}')(\vec{n}\cdot\vec{x}')^m.
\label{eqn:multiexp}
\end{equation} to 1PN order, where $\vec{n}=\vec{x}/r$.  Because of the gauge condition, $\partial_\mu G^{\mu\nu} = 0$, and the retarded potential, the following relations apply,
\begin{equation}
\partial_0 h'_{0k} = n^j \partial_0 h'_{jk},\quad \partial_0 h'_{00} = n^j n^k \partial_0 h'_{jk},
\end{equation} and only need to determine the $h'_{jk}$ components \cite{Will:1993}.  Because, to order, $\partial_\mu J^{\mu\nu} = 0$ by conservation and the source is symmetric, a useful relation is,
\begin{equation}
(\partial^2/\partial t^2)\int \d^3\, x J^{00} x^j x^k = 2\int \d^3x\, J^{jk}
\end{equation}  Therefore,  \ref{eqn:multiexp} becomes,
\begin{equation}
h'_{ij} = -2r^{-1}(\partial^2/\partial t^2)\left(\int \d^3x\, J^{00}(t - r,\vec{x}) x_i x_j\right) + \mathrm{higher\, order}.
\end{equation}  This means that monopole and dipole moments of $J^{ij}$ can be expressed as time derivatives of quadrupole moments of $J^{00}$.  To post-Newtonian order $J^{00} = \rho$, the mass density, and the quadrupole approximation follows.

For the binary system the integral simplifies to,
\begin{equation}
h'_{ij} = \frac{2}{r}\ddot{q}_{jk}(t - r).
\label{eqn:quadpot}
\end{equation} Since from \ref{eqn:grwavepot} the solution is the same as for GR, $h'_{ij}=\bar{h}_{ij}$ and, because $\bar{h}_{ij}$ solves the vacuum equation, $\bar{h}_{ij}=h_{ij}$ obtaining the speed-up formula (\ref{eqn:speedup}) follows as for general relativity:  take a Taylor expansion of \ref{eqn:quadpot} in powers of $r$ to extract the radiation-reaction potential and applying the geodesic equation \ref{eqn:geo}, determine the reaction acceleration of the bodies (\cite{Misner:1973}, Sec. 36.11). Then arrive at \ref{eqn:speedup}. See \cite{Will:1993} and \cite{Misner:1973} for detailed calculations. This proves that the energy loss caused by radiation is the same in GR and the YM theory. The reaction, then, can be modelled by a loss of potential energy between the two bodies.

\subsection{Cosmology}
\label{sec:ae}
If this paper had been written as recently as 15-20 years ago, cosmology could have been addressed more fully within the scope of this paper. In recent years, however, the amount of data available to match to cosmological models has exploded thanks to projects such as WMAP \cite{WMAP:2008}, observations of Type 1a supernovas, and baryon acoustic oscillation measurements. In this section, the Robertson-Walker model is addressed. This is the starting point of the two most prominent numerical approaches, both centered around the Friedmann equations: perturbations of the Einstein field equations and N-body cosmology which models the universe as an N-body Newtonian system.

The Friedmann equations can be derived from Newton's laws \cite{Weinberg:2008}, suggesting that at slow speeds and very weak fields they are valid for the universe as a whole---certainly the present universe is essentially Newtonian at large scales where inhomogeneities are small. In the early universe, however, relativistic speeds and strong gravitational fields predominated and the Newtonian model breaks down for times ``close'' to the Big Bang. 

Cosmological observations have not found major anomalies in the empirical $\Lambda$CDM model, the main cosmological model at this time, but the constraints on alternative theories of gravity are much weaker than those from N-body observations. The Yang-Mills theory predicts some deviations from the standard theory---none of which are challenged by observation because of all the unknowns involved---but it also explains accelerating expansion without a cosmological constant.

One commonly held theory for accelerating expansion is the de Sitter solution to the Einstein Field Equations. In this theory, the accelerating expansion is the result of ordinary expansion over a hyperbolic hypersurface. This is an alternative to the vacuum energy based cosmological constant \cite{Weinberg:2008} but is still necessarily constant because of conservation of energy.

The de Sitter Yang-Mills theory also models the accelerating expansion as expansion over a hyperbolic hypersurface but, instead, the hyperbolic curvature is an active potential that changes with time. Instead of being an {\em ad hoc} solution to the accelerating expansion, the Yang-Mills theory is necessarily built on the de Sitter group from the quantum level up, and accelerating expansion is a natural outgrowth of that symmetry. Although the Yang-Mills theory could be built on the Poincar\'e group for classical predictions, it would not be a semi-simple theory and no longer renormalizable. Hence, the de Sitter symmetry group is in no way {\em ad hoc}. In addition, the two theories, although sharing the de Sitter feature, make different predictions about the past and future of the universe.

\subsubsection{Isotropic, Homogeneous Universe}
Observations of the cosmic microwave background and statistical counts of the distribution of galaxies imply that the universe is statistically isotropic \cite{Wald:1984}.  By the Copernican principle that human beings do not occupy a privileged location in the universe, the universe must also be homogeneous.  An isotropic, homogeneous universe has a simple dynamical description.  For a perfect fluid universe, the stress-energy-momentum tensor in special relativity has the standard form,
\begin{equation}
T^{\mu\nu} = \rho u^\mu u^\nu + p(u^\mu u^\nu + g^{\mu\nu}),
\label{eqn:perfectfluid}
\end{equation} where $\rho$ is mass density and $p$ is pressure \cite{Wald:1984}.  The fundamental Friedmann equations for general relativity are,
\begin{eqnarray}
\dot{a}^2 + k & = & \frac{8\pi \rho a^2}{3},\\
\frac{3\ddot{a}}{a} & = & -8\pi(\rho + 3p),
\end{eqnarray} where $k$ is the curvature constant \cite{Weinberg:2008} and
\begin{equation}
g_{00} = -1,\quad g_{rr} = \frac{a^2(\tau)}{1 - kr^2},\quad g_{\theta\theta} = a^2(\tau)r^2, \quad g_{\phi\phi} = a^2(\tau)r^2\sin^2\theta,
\label{eqn:rwmetric}
\end{equation} is the metric.  Recent observations that the universe's density, the sum of baryon, dark matter, and dark energy densities, is near or at the critical value, $\Omega_b + \Omega_c + \Omega_\Lambda \simeq\Omega_{crit}$, and the universe is spatially flat, $k\approx 0$ \cite{WMAP:2008}\cite{Komatsu:2011}.  Therefore, for the rest of this section, $k=0$.  
From the conservation of energy,
\begin{equation}
\frac{\partial\rho}{\partial \tau} + \frac{3\dot{a}}{a}\left(\rho + p\right) = 0,
\label{eqn:friedenergy}
\end{equation} is a fundamental equation for the evolution of matter and energy.

Turning to the Yang-Mills theory, the equations of motion for the universe are somewhat different and require explanation within the constraints of observation.  Modern cosmology relies on perturbation methods with the Robertson-Walker assumptions of isotropy and homogeneity the zeroth order equations.  Therefore, equations based on these assumptions are critical to any treatment of cosmology.

The isotropic, homogeneous universe under the General Relativity model does not distinguish between matter moving apart within space and space itself moving apart. Hence, redshifts are interpreted as being caused by matter moving apart as space drags it along in its expansion. It is possible, however, to distinguish between the two types of redshift, gravitational and doppler, in ordinary, non-cosmological circumstances such as within the Solar System.

In this section, the theory predicts that, in the Poincar\'e approximation, gravitational redshift expansion is constant in the later universe while Doppler redshift is linear. Thus, in that approximation all observed redshift is caused by the Doppler effect of matter expanding. In the de Sitter theory, however, the influence of the torsion potential, $H$, causes gravitational redshift to increase with time causing total redshift (gravity + doppler) to accelerate with time. This implies that matter is expanding as well as space, causing a dual expansion and, hence, acceleration of the universe's expansion.

The coordinate system is a set of degrees of freedom, that, while having no physical meaning \cite{Norton:2003}, may be used to simplify equations. Comoving coordinates, $g_{00}=-1,g_{ij}=a^2(t)\delta_{ij}$, expand with matter moving {\em within} space. The expansion of space itself is given by the potential, $G_{00} = -1$ and $G_{ij} = b(t)a^2(t)\delta_{ij}$. The torsion potential is non-zero, $H_{\mu\nu\lambda}\neq 0$ as a consequence of the de Sitter Lie algebra, $[V_\mu,V_\nu]\neq 0$, but vanishes in the Poincar\'e approximation where $[V_\mu,V_\nu]= 0$ (see \ref{sec:appendix}). Let $H_{i0i}=-H_{ii0}=c(t)a^2(t)$ and $H_{iij}=-H_{iji}=d(t)a^2(t)$ be the isotropic, homogeneous torsion potential.

 In the comoving coordinates, the conserved current for dust, which is essentially a limit on an N-body system, is $J_{00}=-\rho$ and $J_{ij}=-\rho a^2\delta_{ij}/3$. The spin density is zero by isotropy, $S_{\mu\nu\lambda}=0$. Evaluating the field equations \ref{eqn:motion1}, the time-time equation is,
\begin{equation}
\frac{3\dot{a}(a\dot{b} + b\dot{a} - \dot{a} + ca)}{a^2} = 8\pi \rho,
\label{eqn:rw1}
\end{equation} and the next three (all the same) space-space equations are,
\begin{equation}
\frac{-3\dot{a}\dot{b}a - \ddot{b}a^2 - \ddot{a}ab - \dot{a}^2b + a\ddot{a} - \ddot{b}a^2 - \dot{c}a^2 - 2\dot{a}ac}{a^2} = 8\pi \rho/3.
\label{eqn:rw2}
\end{equation}

From the field equations \ref{eqn:motion2}, the torsion equation for oscillatory torsion is,
\begin{equation}
\ddot{c}a^2 + 3\dot{c}\dot{a}a - 3\dot{a}^2c + \ddot{a}ac - \dot{a}ab + a\dot{a} - \dot{b}a^2 - ca^2 = 0
\end{equation} and the equation for twisting torsion is,
\begin{equation}
\ddot{d}a^2 - \dot{a}^2d + a\dot{a}\dot{d} - a^2b^2d = 0.
\end{equation}

Equation for matter, \ref{eqn:cont2}, simplifies to 
\begin{equation}
\dot{\rho} = -4\frac{\dot{a}}{a}\rho
\label{eqn:matter2}
\end{equation} This does not imply all matter is radiation nor does it support theories such as ``hot'' dark matter. Matter can be slow moving, but its internal energy does affect its behavior in a gravitational field.

The system of three differential equations (\ref{eqn:rw1},\ref{eqn:rw2}, and \ref{eqn:matter2}) govern the gravitational behavior of matter in this universe. The boundary condition are given in terms of present day, $t_0$, and origin, $t=0$, parameters: $a(t_0)=a_0$, $a(0)=0$, $b(0)=0$, $c(0)=0$, and $c(t_0)=c_0$. The assumption of $a(0)=0$ is a classical assumption. Quantum effects take over at $t\ll 1$ and could predict a different initial size of the universe or a quantum ``bounce'' where quantum pressures could reverse a collapsed state.

The equations \ref{eqn:rw1},\ref{eqn:rw2}, and \ref{eqn:matter2} have the non-trivial solution,
\begin{eqnarray}
\label{eqn:scalefactds}
a(t) & = & \frac{a_0}{t_0}t,\\
\rho(t) & = & \rho_0\frac{a_0^4}{a^4},\\
b(t) & = & \frac{1}{12t^2}\left[Ct^4 -3Ct^3 + (12 + 8\pi\rho_0t_0^4)t^2 - 40\pi\rho_0t_0^4t - 32\pi\rho_0t_0^4\right],\\
c(t) & = & \frac{1}{12}\left[-3Ct + 8\pi\rho_0t_0^4\right],
\label{eqn:scalefactdsc}
\end{eqnarray} where $C  =  \frac{1}{3t_0}\left(-12c_0 + 8\pi\rho_0t_0^4\right)$. The equation for $d(t)$ is not expressible in closed form but is the solution to a second order ODE. Note: these results were obtained with Maple\textsuperscript{TM}'s tensor package and PDE solver.

The scale factor, $a(t)$, conforms to a universe with a Minkowski metric sometimes called the Milne universe or simply the linear model \cite{Liebscher:2005}.  The Milne cosmological model has recently seen a resurgence as an alternative to the standard $\Lambda$CDM model and has some compelling features:  It gives the same age for the universe as the $\Lambda$CDM model \cite{Kutschera:2007}.  Nucleosynthesis has been fit to observations for the linear model in \cite{Sethi:1999} with some problems that I discuss below \cite{Kaplinghat:2000}.  Predictions for Type 1a supernovae are very close to those of the standard model.  The Milne universe has angular distances in the CMB of about $1.1$ degrees, also close to the observed value \cite{Benoit:2008a} \cite{Benoit:2008b}.

Although the rate of matter expansion is linear, the gravitational potential, $b(t)$, shows an acceleration in total expansion rate (space + matter), related to $G_{ij}\delta^{ij}/3=b(t)a^2(t)\sim O(t^4)$, meaning that the redshift of distant galaxies should be accelerating linearly with time. It is clear that the theory deviates from the linear or Milne theory (against which there have been objections based on Big Bang Nucleosynthesis \cite{Kaplinghat:2000}) in relation to how strongly the universe deviates from the Poincar\'e approximation. At the present time, the acceleration is not well understood enough to determine whether it is linear, exponential (as in models with a cosmological constant), or some other function of time. Hence, should better measurements become available that show the rate of acceleration, they may test the theory's predictions.

\subsubsection{The Poincar\'e approximation}

If the hypersurface over which particles translate in the de Sitter model has infinite size, translations become straight, and the Poincar\'e group is the result. When the Poincar\'e approximation is made such that we change the equations to conform to the Lie algebra with $[V^\mu,V^\nu] \approx 0$ (see \ref{sec:lorentz}), the expansion no longer accelerates:

\begin{eqnarray}
\label{eqn:scalefact}
a(t) & = & a_0 t/t_0,\\
\rho(t) & = & \rho_0\left(\frac{a_0}{a}\right)^4,\\
b(t) & = & 1 - \frac{8\pi \rho_0}{3(a_0^4/t_0^4)t^2} + \frac{K}{(a_0/t_0) t},
\label{eqn:scalefactb}
\end{eqnarray} where $K = (b_0 + 8\pi\rho_0/(3a_0^4/t_0^2) - 1)a_0$ and $b_0 = b(t_0)$. Note: $b(t)\rightarrow 1$ as $t\rightarrow \infty$ indicating that the Poincar\'e universe has no accelerating expansion at its present age. The de Sitter Lie algebra offers an explanation: when $[V^\mu,V^\nu] = iM^{\mu\nu}$ (see \ref{sec:lorentz}), momentum, related to the $V^\mu$ generators, produces spin. Because $[M_{\mu\nu},P_\rho]  =  i(\eta_{\mu\rho}V_\nu - \eta_{\nu\rho} V_\mu)$ spin gravitons couple to momentum gravitons to create more momentum gravitons. Since the potential for momentum gravitons is $G_{\mu\nu}$, which relates to the expansion of space, this relationship causes accelerating redshift.

\subsubsection{Accelerating Expansion}

\begin{figure}[h]
\centering
\includegraphics[width = 0.6\textwidth]{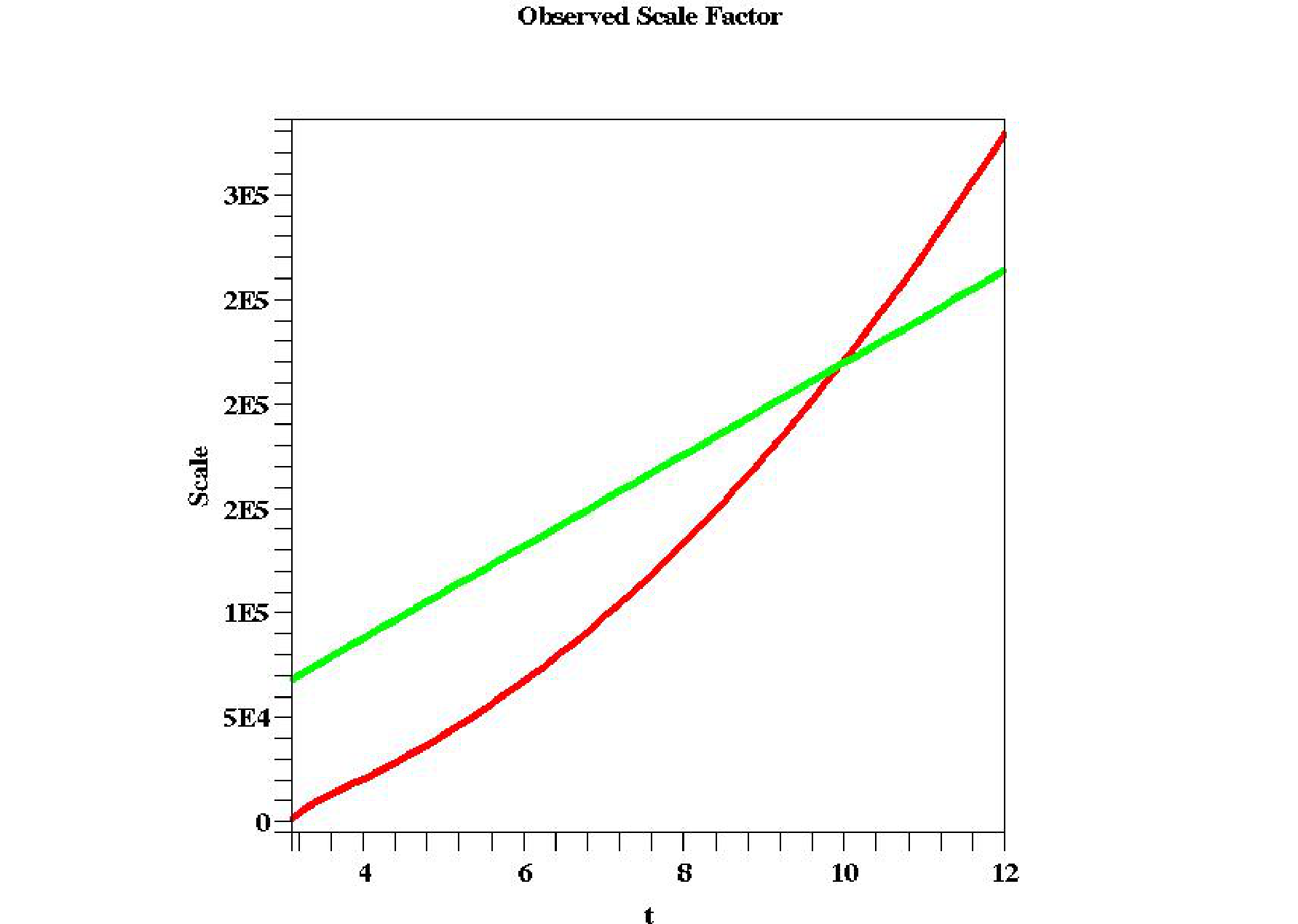}
\caption{The $\textrm{Observed Scale Factor} = \sqrt{a^2b}$ has very different behavior in the de Sitter vs. Poincar\'e approximation. The straight line is given by the Poincar\'e equations \ref{eqn:scalefact}-\ref{eqn:scalefactb} and has no acceleration. The curving line is given by \ref{eqn:scalefactds}-\ref{eqn:scalefactdsc} and shows acceleration. Boundary conditions have been chosen such that the two lines cross at approximately $t=t_0$: $a_0=100,t_0=10,c_0=1,\rho_0=1$ for the de Sitter graph and $b_0=1,\rho_0=1,a_0=220,000,t_0=10$ for the Poincar\'e graph. Note that the Poincar\'e universe is much older.}
\label{fig:ae} 
\end{figure}

Like the de Sitter solution to the Einstein Field equations, the de Sitter Yang-Mills theory explains the accelerating expansion. In the Poincar\'e group approximation, neither phenomena is predicted. The expansion of the universe in the Poincar\'e and de Sitter models is depicted in Fig. \ref{fig:ae}. Only the de Sitter model shows acceleration at the present day.

Although at first glance it may seem that by tweaking the initial conditions, the prediction will vanish, it is, in fact, robust ``end-point'' behavior and emerges as the scale factor becomes very large.

\section{Potential Disagreements}
\label{sec:disagreement}
Observations of gravity are classed into a four basic types: (1) Newtonian gravity which applies to very weak fields and slow velocities that pertain to most galaxies, stars, and orbits of the outer planets, (2) weak field gravity responsible for classic effects such as redshift, light bending, time dilation, and perihelion precession, (3) strong field gravity responsible for binary pulsar precession, orbital damping, and gravitational waves, and (4) cosmology where assumptions such as homogeneity and isotropy greatly simplify equations as well as perturbative cosmology (not addressed in this paper).  Included are tests of the Strong Equivalence Principle (SEP).

Going through each of the four types of tests:

{\em Newtonian gravity.} Both theories subsume Newtonian gravity at very weak field strengths and slow velocities, easily demonstrated from the field equations. There is no disagreement here which is not true of all alternatives to GR. Conformal gravity is an example of a theory that diverges from Newtonian gravity \cite{Mannheim:2006}.

{\em Weak Fields.} As shown in section \ref{sec:traj}, both theories agree with classic effects up to linear order in the Schwarzschild radius over the distance.  Measurements of redshift, light bending, perihelion precession, and time dilation have not been made to quadratic accuracy in Schwarzschild coordinates (as opposed to isotropic coordinates where the coordinate change introduces artifact quadratic term); therefore, none of these experiments contradict either theory.

{\em Strong Fields.} As \cite{Kramer:2006} mentions, measurements of binary pulsar precession, while some of the most precise measurements of relativistic gravity ever made, are not sufficiently accurate to confirm beyond the 1PN equations of motion to the order given in Sec. \ref{sec:1pn}.  Orbital damping caused by radiation reaction is not confirmed beyond the quadrupole approximation, and neither GR nor the YM theory predict a dipole moment.  Therefore, as shown in section \ref{sec:1pn}, none of these measurements contradicts either theory nor any other measurements of multibody systems.  The YM theory's geodesic equation \ref{eqn:geo} is also the same as in relativity; therefore, since SEP is satisfied (Sec. \ref{sec:sep}) and compact self-gravitating bodies behave as if they are test bodies, it agrees with the 1PN equations of motion of general relativity.  Gravitational wave measurements promise to provide higher order estimates which may show a violation of one of the theories, but these, as yet, are not available. Few alternatives to GR satisfy SEP and some have been ruled out by N-body observations that have constrained the PPN parameters.

{\em Cosmology.} Several sources including Type Ia supernovae (SN), baryon acoustic oscillation (BAO), and the cosmic microwave background (CMB) provide data that can be used to constrain various theories \cite{Komatsu:2011}.  The Robertson-Walker model is essentially a toy or zeroth order model because it assumes no fluctuations or deviations from perfect uniformity and isotropy in the universe.  First order perturbation of the field equations about the Robertson-Walker solution combined with numerical solvers can indicate whether observed fluctuations in the CMB agree with the theory, and this is an important future direction of research for the YM theory.  

Studies of Big Bang Nucleosynthesis (BBN) where the predicted quantities of light elements such as Helium, Lithium, and Deuterium are compared with predicted quantities can also constrain some theories and several studies of Tensor-Scalar theory have been done \cite{Damour:1999}\cite{Coc:2006}.  These studies tend to focus on the effects of the scalar field on nucleosynthesis and are not directly applicable to the YM theory which has no scalar field.

While the linear model explains features of the later universe as well as the standard $\Lambda$CDM model \cite{Benoit:2008a}, a single power law scale factor has proved insufficient to explain the early universe where the power is sharply constrained to around $a\sim t^{0.55}$ \cite{Kaplinghat:2000} unless additional modes of production of deuterium are invoked in the later universe. Because the YM theory predicts a multi-term polynomial gravitational redshift evolution on top of the Doppler redshift evolution, however, the theory has a (at least) dual epoch model. Because the present values of $b(t)$ and $c(t)$ are unknown, however, whether the theory is able to predict abundances of light elements is left for future work.

\section{Future Tests}
\label{sec:novel}
General Relativity and the Yang-Mills theory agree for the linearized Schwarzschild metric predictions and 1PN equations of motion, but they fundamentally diverge at higher order motion.  The 2PN order should be measurable for binary pulsars within the next few decades as additional data is collected about known systems (the more data collected the smaller the error bound). It is doubtful that the 2nd order of the Schwarzschild geometery will be measurable within the Solar system any time soon because the required accuracy would have to be thousands of times better than current, already extremely sensitive, instruments provide, at which point significant and potentially unknown variations in Earth's or the Sun's density would make it difficult to predict the measurements. Gravitational wave detectors such as LIGO and VIRGO may also be capable of detecting much higher order effects (well into the quadratic or even cubic realm) from inspiraling black holes and other closely spaced, highly massive objects. Because general relativity and the Yang-Mills theory no longer agree at this level, these observations will be able to rule out one of the theories. Another important test is a quantitative measurement of the values of $b_0$ and $c_0$ from Sec. \ref{sec:ae} which may allow a quantitative prediction of the rate of acceleration of the universe, an important prediction of the theory.

Quantum predictions are significant because the theory, in order to be renormalizable, requires mass to be quantized so that the coupling constant has zero mass dimension.  Without that assumption, the coupling constant would have negative mass dimension of -2 and the theory would not be renormalizable \cite{Zee:2003}.

\section{Related Work}
\label{sec:relwork}
While this paper has focused on the problems with the prevailing theory of gravity, general relativity, several other theories have been proposed to solve quantum gravity and explain macroscopic observations.  The most important empirical theory is the Lambda-CDM model which combines the Einstein field equations with a cosmological constant and cold dark matter.  The field equations \ref{eqn:motion1}-\ref{eqn:motion2} do away with the cosmological constant and challenge some of the assumptions of the model, but the YM theory does not challenge the prevailing theory of dark matter.  

The most prominent theory of quantum gravity is string theory and its derivative M-theories which, rather than being a simple theory of gravity, is an extensive modification of physical law positing that all matter is composed of strings, attempting to unify all forces \cite{Zee:2003}\cite{Rovelli:2004}.  Although unification is not its purpose, in representing gravity identically to the other three forces, the theory can be unified with them at a high enough energy without requiring any significant additional physical assumptions such as strings or additional spacetime dimensions. 

The next most significant model of quantum gravity is loop quantum gravity \cite{Rovelli:2004}.  Loop quantum gravity has a different approach from that given in this paper.  In making the assumption that gravity has only diffeomorphism covariance and is dominated by a metric geometry, it retains the symmetry of general relativity and adds to its complexity by introducing spin foams, i.e. discretizations of spacetime, in order to avoid blow-ups of the quantum variables.  It currently makes no predictions that are testable and, furthermore, has not been shown to agree with classical general relativity in its entirety.

\section{Conclusion}
I have shown that the Yang-Mills theory agrees with all of general relativity's predictions for N-body motion, while I have given strong evidence that it can also do so on the cosmological stage. Thus, it is not possible to say that Yang-Mills theory cannot replace general relativity given current data. Further data is needed to make that determination. The Yang-Mills theory, however, is compelling because, unlike general relativity, it is a renormalizable quantum theory of gravity. In addition, it does not couple to vacuum energy, an unobserved prediction of general relativity. Unlike leading theories, it also requires no significant additional physical assumptions other than the de Sitter group symmetry in a Yang-Mills formalism---a formalism that is well established as explaining all other forces.

The most important prediction beyond general relativity, and what makes the Yang-Mills theory even more interesting besides its renormalizability, is the prediction of the accelerating expansion of the universe from the de Sitter group. While the de Sitter solution to the Einstein Field equations is an {\em ad hoc} solution to accelerating expansion, the Yang-Mills theory is built on the de Sitter group from the ground up and thus explains accelerating expansion as an outgrowth of quantum symmetry. Although qualitative, it has the potential to be a quantitative prediction with additional data to constrain the current conditions of the universe within the context of this theory, particularly values of $b_0$ and $c_0$. Further confirmations may be obtained from detection of gravitational waves, observations of supermassive objects such as black holes, and additional cosmological measurements.

\appendix

\section{The de Sitter Group Lie Algebra}
\label{sec:lorentz}
In addition to rotations and boosts, all physical theories are translation invariant.  This confirms the assumption that there is no ``special'' place in the universe.  By Noether's theorem, translation invariance causes momentum to be conserved (including energy).  A translation by an amount $v_\mu$ is achieved by a 5$\times$5 matrix, $T=v^\mu V_\mu$ such that $V_\mu = (V_t, V_i)$ are four generators.  Given a 4-vector $u_\mu$ if $w = (u_\mu, 1)$, then $w' = T w = (u_\mu + v_\mu, 1)$.  It is well known, however, that prior to the introduction of Lorentz covariance, mechanical theories such as Newtonian gravity were Galilean invariant, $R^3\otimes$SO(3) rather than Lorentz invariant, SO(3,1).  A reasonable extension of the Poincar\'e group, then, is from $R^{(3,1)}\otimes$SO(3,1) to SO(4,1) assuming that the radius of curvature in the fifth dimension is large so that it appears to be Poincar\'e invariant for small rotations.

In the de Sitter group, $x_4$ is a fourth spatial dimension and has ordinary rotations with respect to the other three spatial dimensions and boosts with respect to time.  The anti-de Sitter group has $x_4$ as a second time dimension.  Let $V_\mu$ be rotations/boosts in the $x_\mu$-$x_4$ plane.

The Poincar\'e group has the covariant Lie algebra,
\begin{eqnarray}
\,[V_\mu,V_\nu] & = & 0\\
\,[M_{\mu\nu},V_\rho] & = & i(\eta_{\mu\rho}V_\nu - \eta_{\nu\rho} V_\mu)\\
\,[M_{\mu\nu},M_{\rho\sigma}] & = & i(\eta_{\mu\rho}M_{\nu\sigma} - \eta_{\mu\sigma}M_{\nu\rho} - \eta_{\nu\rho} M_{\mu\sigma} + \eta_{\nu\sigma}M_{\mu\rho}),
\end{eqnarray} for $\eta_{\mu\nu}$ the Minkowski metric where $U=\exp[\frac{i}{2}\omega_{\mu\nu}M^{\mu\nu}]$ is a Lorentz operation and $U=\exp[ia_\mu V^\mu]$ is the translation operation \cite{Ohlsson:2011}.  The only alteration that the de Sitter group makes is that the ``translation'' operators now generate a Lorentz rotation or boost,
\begin{equation}
\,[V_\mu,V_\nu] = iM_{\mu\nu},
\end{equation} in which case 
\begin{equation}
U = \exp[\frac{i}{2}\omega_{\mu\nu}M^{\mu\nu} + ia_\mu V^\mu],\nonumber
\end{equation} is the de Sitter operation \cite{Aldrovandi:1995}.

\section{Derivation of the 1PN potential}
\label{sec:appendix}
The potential at a spacetime point $(t,{\vec x})$ in the given inertial frame can be solved by the integral equations \cite{Will:2006}, $h_{\mu\nu}=(G_{\mu\nu} - \eta_{\mu\nu})/2$:
\begin{eqnarray}
\label{eqn:integral}
h_{\mu\nu}(t,{\vec x}) & = & -\int_{\mathcal{C}} \d^3 x' \frac{J_{\mu\nu}(t - |{\vec x} - {\vec x}'|,{\vec x}')}{|{\vec x} - {\vec x}'|},
\end{eqnarray} where $\mathcal{C}$ is the past-light cone and bodies have small polarized acceleration and angular momentum is small.

The integral equation may be solved by iteration in which a trial potential $h_{\mu\nu}=0$ is inserted into the integrand and the integral solved to arrive at a new potential $h'_{\mu\nu}$ which is then reinserted to achieve a potential $h''_{\mu\nu}$ and so on until the required accuracy is achieved.  
In order to derive the 1PN equations of motion general relativity requires two iterations but here we only require one, meaning that the 1PN equations for this theory are linear.

Let the baryon density be $\rho_0$, and the Newtonian potential is given by,
\begin{equation}
U = \int \d^3x'\, \frac{\rho_0(t,\vec{x}')}{|\vec{x} - \vec{x}'|}.
\end{equation}  Let $U\sim \epsilon^2$ and $v/c\sim \epsilon$ where $v$ is the average velocity and $c$ is the speed of light in vacuum.  The first post-Newtonian corrections require that $h_{00}$ be given up through order $\epsilon^4$, $h_{0j}$ through order $\epsilon^3$ and $h_{ij}$ through $\epsilon^2$, correcting the Newtonian order,
\begin{equation}
h_{00} = U,\, h_{0j} = 0,\, h_{ij}=0.
\end{equation}

The G-field derives from the equations of Sec. \ref{sec:yangmills} above and the stress-energy-momentum tensor for baryon dust.  Each baryon particle or compact spherical mass (such as a neutron star or black hole) $p$ has a stress-energy-momentum tensor in its rest frame in a spherical coordinate system with the particle at the origin,
\begin{equation}
J_{00} = m_p\delta^3(\vec{x}),\, J_{rr} = m_p\delta^3(\vec{x}).
\end{equation}  The latter is related to the potential $h_{rr}$ which has a transformation, 
\begin{equation}
h_{ij} = h_{rr}\frac{x_i x_j}{|\vec{x}|^2},
\end{equation} which is
\begin{equation}
h_{ij} = m_p\frac{(x_i - x_i')(x_j - x_j')}{|\vec{x} - \vec{x}'|^3}
\end{equation}  for a particle at $\vec{x}'$.  To obtain the potential for all the dust, sum all the potentials for all the baryons by superposition of the potentials,
\begin{equation}
\tilde{h}_{ij} = \int \d^3x'\, \frac{\rho_0(t,\vec{x}')(x_i - x_i')(x_j - x_j')}{|\vec{x} - \vec{x}'|^3} + O(\epsilon^4).
\end{equation}  By a non-Abelian gauge transformation to order \cite{Misner:1973},
\begin{equation}
h_{ij} = \tilde{G}_{ij} + \partial_i \theta_j,
\label{eqn:gaugetrans}
\end{equation} where $\theta_\mu = \partial_\mu \chi$ and
\begin{equation}
\chi = \int \d^3x'\, \rho_0(t,\vec{x}')|\vec{x}-\vec{x}'|,
\end{equation} we have
\begin{equation}
h_{ij} = \delta_{ij}U + O(\epsilon^4).
\label{eqn:1pngij}
\end{equation}

The next two parts of the G-field, $h_{0j}$ and $h_{00}$, require particle velocities and time dilation to be included.  The current tensors, $J_{\mu\nu}(p)$ of each baryon $p$, are in the rest frame, but, because the bodies are moving with respect to the observer at infinity and subject to the gravitational fields of one another, the tensors must be (1) boosted from comoving frame $\bar{x}^\mu$ to the observer's frame, $x^\mu$, and (2) subject to a non-Abelian gauge transformation from the zero gravitational field (free fall) in the rest frame of the body to the gravitational field in the rest frame of the observer,
\begin{equation}
J_{\mu\nu}(p) = \frac{d \bar{x}^\mu}{dx^\alpha}\frac{d \bar{x}^\nu}{dx^\beta} J_{\alpha\beta}(p) + h_{\mu\nu}m_p\delta^3(\vec{x}-\vec{x}_p),
\end{equation} where $\Lambda^\mu{}_\nu = d\bar{x}^\mu/dx^\nu$ is a Lorentz transformation matrix for a boost density $\vec{v}(t,\vec{x})$ ($c=1$)\cite{Misner:1973} and $h_{\mu\nu}$ is the G-field not including the contribution of the mass at that point.  These two transformations are equivalent to the two step, boost-and-coordinate-transform method of general relativity \cite{Misner:1973} which reflects the effects of both velocity and gravitational fields on measurements of stress-energy-momentum tensors.

For a baryon $p$ with mass $m_p$ at $\vec{x}'$ with a rest frame potential $\bar{h}_{\mu\nu}$ and velocity $\vec{v}$ ($c=1$),
\begin{equation}
h_{0j} = \bar{h}_{00}\frac{\partial \bar{x}^0}{\partial x^0}\frac{\partial \bar{x}^0}{\partial x^j} + \bar{h}_{ij}\frac{\partial \bar{x}^i}{\partial x^0}\frac{\partial \bar{x}^j}{\partial x^j},
\end{equation} where
\begin{equation}
\bar{h}_{ij} = \frac{m_p(x_i - x_i')(x_j - x_j')}{|\vec{x} - \vec{x}'|^3}.
\end{equation}  To order this becomes,
\begin{equation}
\tilde{h}_{0j} = -\frac{m_p}{|\vec{x} - \vec{x}'|}\left(v_j + \frac{[(\vec{x} - \vec{x}')\cdot \vec{v}](x_j - x_j')}{|\vec{x} - \vec{x}'|^2}\right) + O(\epsilon^5).
\end{equation}
Under the same gauge transformation as above,
\begin{equation}
h_{0j} = \tilde{h}_{0j} - \partial_0 \theta_j,
\end{equation} and summing over all baryons, the potential is,
\begin{equation}
h_{0j} = -2V_j + O(\epsilon^5),
\label{eqn:1png0j}
\end{equation} where
\begin{equation}
V_j = \int \d^3x'\, \frac{\rho_0(t,\vec{x}')v_j(t,\vec{x}')}{|\vec{x} - \vec{x}'|}.
\end{equation}

The time-time field is,
\begin{equation}
\tilde{h}_{00} = U + 2\Psi + \Phi + O(\epsilon^6),
\end{equation} where
\begin{equation}
\Psi(t,\vec{x}) = \int \d^3x'\, \frac{\rho_0(t,\vec{x}')(v^2 + U)}{|\vec{x} - \vec{x}'|},
\end{equation} and
\begin{equation}
\Phi(t,\vec{x}) = \int \d^3 x'\, \frac{\rho_0(t,\vec{x}')}{|\vec{x} - \vec{x}'|}\left\{\frac{[(\vec{x} - \vec{x}')\cdot \vec{v}]^2}{|\vec{x} - \vec{x}'|^2} - v^2\right\}.
\end{equation}  The gauge transformation, $h_{00} = \tilde{G}_{00} + \partial_0^2\chi$ (which we have already applied to the other potentials and hence must carry over to this one), eliminates the last term, and gives
\begin{equation}
h_{00} = U + 2\Psi + O(\epsilon^6).
\label{eqn:1png00}
\end{equation}  Missing is a term of order $U^2$ that appears in metric PPN formalism which indicates the nonlinearity in the superposition of the potential.  In GR, however, this term is an artifact of the choice of isotropic coordinates and vanishes with a change of coordinates.

 In general relativity, the change of coordinates, $r=\bar{r}(1 + M/2\bar{r})^2$, switches the Schwarzschild solution to isotropic coordinates \cite{Misner:1973},
\begin{equation}
ds^2 = -\left(\frac{1 - M/2\bar{r}}{1 + M/2\bar{r}}\right)^2 dt^2 + (1 + M/2\bar{r})^4[d\bar{r}^2 + \bar{r}^2(d\theta^2 + \sin^2\theta d\phi^2)].
\end{equation}  Because the change in coordinates, however, depends on the potential itself, this introduces a nonlinearity that previously did not exist in the solution.  Like the Schwarzschild solution in Schwarzschild coordinates, the Yang-Mills theory 1PN solution does not contain any nonlinearity because it is not in this nonlinear coordinate system.  The Yang-Mills spherically symmetric solution given in spherical coordinates,
\begin{equation}
h_{00} = h_{rr} = M/r,
\end{equation} under the same change of variables, becomes to 1PN order,
\begin{equation}
h_{00} = M/\bar{r} - 2(M/\bar{r})^2 + O(\epsilon^6),
\end{equation} 
\begin{equation}
h_{0i} = h_{i0} = O(\epsilon^5)
\end{equation} and
\begin{equation}
h_{ij} = \delta_{ij}M/\bar{r} + O(\epsilon^4).
\end{equation}

While one could make a change of the flat spacetime coordinates to bring the YM potential in line with the GR metric, there is no purpose in doing so.  Since both GR and the Yang-Mills theory are generally covariant, changing coordinates does not change predictions.  It does, however, require that observations be input in the appropriate coordinate system since the Schwarzschild $r$ is not the same as the 1PN $\bar{r}$.  In General Relativity, the Schwarzschild $r$ is simply a coordinate not a distance.  Meanwhile, because of the potential dependent metric in ``isotropic'' coordinates, $\bar{r}$ is just a coordinate in both theories and does not refer to distance directly.

The remaining equations (which can be derived from the field equations) are the equation of motion. The continuity equations (\ref{eqn:cont}) to 1PN order are the four equations,
\begin{equation}
\partial_\nu J^{\nu\alpha} = 0,
\end{equation} each of which is the same as the electromagnetic continuity equation.  Linearized general relativity has the same problem \cite{Wald:1984} and care needs to be taken that the nonlinear, gauge covariant equations of motion (\ref{eqn:cont}) be used instead.  For $N$-body problems, which are generally the ones that gravitation deals with, however, the geodesic equation is sufficient.  This is because the Yang-Mills theory satisfies the Strong Equivalence Principle (SEP) and internal structure does not affect the motion of bodies.  Therefore, any spherical, compact body moves as a test particle.
 
\bibliography{sg2}

\begin{thebibliography}{10}

\bibitem{Zwicky:1937}
F.~{Zwicky}.
\newblock On the masses of nebulae and of clusters of nebulae.
\newblock {\em Astrophysical J.}, 86:217, 1937.

\bibitem{Rubin:1970}
V.~{Rubin}.
\newblock Rotation of the andromeda nebula from a spectroscopic survey of
  emission regions.
\newblock {\em Astrophysical J.}, 159:379, 1970.

\bibitem{Clowe:2006}
D.~{Clowe {\em et al.}}
\newblock A direct empirical proof of the existence of dark matter.
\newblock {\em Astrophysical Journal Letters}, 648:109--114, 2006.

\bibitem{WMAP:2008}
NASA.
\newblock Wmap mission results, 2008.
\newblock Updated: March 7, 2008. Accessed: March 18, 2009. {\texttt
  http://map.gsfc.nasa.gov/news}.

\bibitem{Zee:2003}
A.~{Zee}.
\newblock {\em Quantum Field Theory in a Nutshell}.
\newblock Princeton UP, Princeton, 2003.

\bibitem{Wald:1984}
R.~{Wald}.
\newblock {\em General Relativity}.
\newblock U. Chicago UP, Chicago, 1984.

\bibitem{Misner:1973}
C.~W. {Misner}, K.~S. {Thorne}, and J.~A. {Wheeler}.
\newblock {\em Gravitation}.
\newblock W. H. Freeman, San Francisco, 1973.

\bibitem{Kramer:2006}
M.~{Kramer} and I.~H. {Stairs {\em et al.}}
\newblock Tests of general relativity from timing double pulsar.
\newblock {\em Science}, 314(5796):97--102, 2006.

\bibitem{Komatsu:2011}
E.~{Komatsu {\em et al.}}
\newblock Seven-year wilkinson microwave anisotropy probe (wmap) observations:
  Cosmological interpretation.
\newblock {\em The Astrophysical Journal Supplement Series}, 192(2):18, 2011.

\bibitem{Weinberg:2008}
S.~{Weinberg}.
\newblock {\em Cosmology}.
\newblock Oxford UP, Oxford, UK, 2008.

\bibitem{Will:1993}
C.~{Will}.
\newblock {\em Theory and experiment in gravitational physics}.
\newblock Cambridge UP, Cambridge, UK, 1993.

\bibitem{Riegert:1986}
R.~J. {Riegert}.
\newblock {\em Classical and Quantum Conformal Gravity}.
\newblock PhD thesis, UCSD, 1986.

\bibitem{Mannheim:2006}
Philip~D. Mannheim.
\newblock Alternatives to dark matter and dark energy.
\newblock {\em Progress in Particle and Nuclear Physics}, 56(2):340 -- 445,
  2006.

\bibitem{Freeman:2006}
K.~{Freeman} and G.~{McNamara}.
\newblock {\em In search of dark matter}.
\newblock Springer, 2006.

\bibitem{Wise:2010}
Derek~K Wise.
\newblock Macdowell–mansouri gravity and cartan geometry.
\newblock {\em Classical and Quantum Gravity}, 27(15):155010, 2010.

\bibitem{Wesson:2003}
P.~S. {Wesson}.
\newblock Is mass quantized?
\newblock {\em Mod. Phys. Lett. A}, 19:1995--2000.

\bibitem{Parikh:2005}
M.~K. {Parikh} and E.~P. {Verlinde}.
\newblock De sitter space with finitely many states: A toy story.
\newblock In {\em Proc. of 10th Marcel Grossman Meeting on GR (Rio de Janeiro,
  Brazil}, page 2346. World Sci., 2005.

\bibitem{Benoit:2008a}
A.~{Benoit-Levy} and G.~{Chardin}.
\newblock Observational constraints of a milne universe.
\newblock In {\em Proc. of the 43rd Rencontres de Moriond}, 2008.

\bibitem{Kaplinghat:2000}
M.~{Kaplinghat}, G.~{Steigman}, and T.~P. {Walker}.
\newblock Nucleosynthesis in power-law cosmologies.
\newblock {\em Phys. Rev. D}, 61(103507), 2000.

\bibitem{Jackson:1999}
J.~D. {Jackson}.
\newblock {\em Classical Electrodynamics}.
\newblock Wiley, Hoboken, NJ, 1999.

\bibitem{Norton:2003}
J.~{Norton}.
\newblock {\em General covariance, gauge theories and the Kretschmann
  objection}.
\newblock Cambridge UP, Cambridge, UK, 2003.

\bibitem{Joshi:1985}
D.~C. {Joshi}, M.~P. {Benjwal}, and S.~{Kumar}.
\newblock Classical yang mills theory in presence of electric and magnetic
  charges.
\newblock {\em Acta Physica Polonica}, B16(10):901--908, 1985.

\bibitem{Will:2006}
C.~{Will}.
\newblock The confrontation between general relativity and experiment.
\newblock {\em Living Review: Relativity}, 9, 2006.

\bibitem{tHooft:2005}
Gerardus 't~Hooft.
\newblock {\em 50 Years of Yang-Mills theory}.
\newblock World Scientific, Singapore, 2005.

\bibitem{Muller:2010}
H.~{M\"uller}, A.~{Peters}, and S.~{Chu}.
\newblock A precision measurement of the gravitational redshift by the
  interference of matter waves.
\newblock {\em Nature}, 463:926, 2010.

\bibitem{Everitt:2011}
C.~W.~F. {Everitt {\em et al.}}
\newblock Gravity probe b: Final results of a space experiment to test general
  relativity.
\newblock {\em Phys. Rev. Lett.}, 106:221101, May 2011.

\bibitem{Einstein:1918}
A.~{Einstein}.
\newblock {\em Preuss. Akad. Wiss. Berlin, Sitzber}, 154, 1918.

\bibitem{Peters:1963}
P.~C. {Peters} and J.~{Matthews}.
\newblock Gravitational rational from point masses in a keplerian orbit.
\newblock {\em Phys. Rev.}, 131(1):435--440, 1963.

\bibitem{Weisberg:2005}
J.~M. {Weisberg} and J.~H. {Taylor}.
\newblock Relativistic binary pulsar b1913+16: Thirty years of observations and
  analysis.
\newblock {\em ASP Conf. Series}, 328:25--31, 2005.

\bibitem{Thorne:1980}
K.~S. {Thorne}.
\newblock Multipole expansions of gravitational radiation.
\newblock {\em Rev. Mod. Phys.}, 52:299--324, 1980.

\bibitem{Liebscher:2005}
D-E {Liebscher}.
\newblock {\em Cosmology}.
\newblock Springer, Heidelberg, 2005.

\bibitem{Kutschera:2007}
M.~{Kutschera} and M.~{Dyrda}.
\newblock Coincidence of age in $\lambda$cdm and milne cosmologies.
\newblock {\em Acta Phys. Polonica B}, 38(1):215--217, 2007.

\bibitem{Sethi:1999}
M.~{Sethi}, A.~{Batra}, and D.~{Lohiya}.
\newblock Comment on ``observational constraints on power-law cosmologies''.
\newblock {\em Phys. Rev. D}, 60(108301), 1999.

\bibitem{Benoit:2008b}
A.~{Benoit-Levy} and G.~{Chardin}.
\newblock A symmetric milne universe: a second concordant universe?
\newblock In {\em SF2A 2008}, 2008.

\bibitem{Damour:1999}
T.~{Damour} and B~{Pichon}.
\newblock Big bang nucleosynthesis and tensor-scalar gravity.
\newblock {\em Phys. Rev D}, 59(123502), 1999.

\bibitem{Coc:2006}
A.~{Coc}, K.~{Olive}, J-P. {Uzan}, and E.~{Vangioni}.
\newblock Big bang nucleosynthesis constraints on scalar-tensor theories of
  gravity.
\newblock {\em Phys. Rev. D}, 73(083525), 2006.

\bibitem{Rovelli:2004}
C.~{Rovelli}.
\newblock {\em Quantum Gravity}.
\newblock Cambridge UP, Cambridge, UK, 2004.

\bibitem{Ohlsson:2011}
T.~{Ohlsson}.
\newblock {\em Relativistic Quantum Physics: From Advanced Quantum Mechanics to
  Introductory Quantum Field Theory}.
\newblock Cambridge UP, 2011.

\bibitem{Aldrovandi:1995}
R.~Aldrovandi and J.~G. Pereira.
\newblock {\em {An Introduction to Geometrical Physics}}.
\newblock World Scientific, 1995.

\end{thebibliography}

\end{document}